\documentclass[%
 reprint,
showpacs,
 amsmath,amssymb,
 aps,
 pra,
floatfix,
superscriptaddress
]{revtex4-1}
\usepackage[utf8]{inputenc}
\usepackage[T1]{fontenc}
\usepackage[]{graphicx}
\usepackage{color}
\usepackage[dvipsnames]{xcolor}
\usepackage[english]{babel}
\usepackage[normalem]{ulem}
\usepackage[bookmarks,bookmarksopen,bookmarksdepth=2]{hyperref}
\hypersetup{
    colorlinks=true,
    citecolor=blue,
    linkcolor=blue,
    filecolor=magenta,
    urlcolor=cyan,
}

\usepackage{microtype}
\usepackage{braket}
\usepackage{esvect}
\usepackage[export]{adjustbox}
\usepackage{dsfont}
\usepackage[T1]{fontenc}
\usepackage[utf8]{inputenc}
\usepackage{url}
\usepackage{orcidlink}

\usepackage{amsmath,amsfonts,amssymb,amsthm}
\usepackage{mathtools}
\usepackage{bbm}
\usepackage{mathrsfs,physics,xfrac,cancel,tensor,relsize,tikz,lipsum,quantikz}

\newcommand*{\tildehat}[1]
{
\hat{\rule{0ex}{-1.2ex}\mkern-3mu \widetilde{#1}}
}

\newcommand{\expv}[1]{\langle#1\rangle}
\newcommand{\col}[2]{\|#1\rangle\langle#2\|}

\newcommand{\party}[2]{\ensuremath{\frac{\partial{#1}}{\partial{#2}}}}

\begin{document}

\title{{Modeling local decoherence of a spin ensemble using a generalized Holstein-Primakoff mapping to a bosonic mode}}
\author{Andrew Kolmer Forbes\orcidlink{0009-0003-8730-8007}}
\email{aforbes@unm.edu}
\author{Philip Daniel Blocher\orcidlink{0000-0001-6302-7567}}
\author{Ivan H. Deutsch}
\affiliation{Center for Quantum Information and Control, Department of Physics and Astronomy, University of New Mexico, Albuquerque, New Mexico 87131, USA}

\begin{abstract}
We show how the decoherence that occurs in an entangling atomic spin-light interface can be simply modeled as the dynamics of a bosonic mode. Although one seeks to control the collective spin of the atomic system in the permutationally invariant (symmetric) subspace, diffuse scattering and optical pumping are local, making an exact description of the many-body state intractable. To overcome this issue we develop a generalized Holstein-Primakoff approximation for collective states which is valid when decoherence is uniform across a large atomic ensemble. In different applications the dynamics is conveniently treated as a Wigner function evolving according to a thermalizing diffusion equation, or by a Fokker-Planck equation for a bosonic mode decaying in a zero temperature reservoir.  We use our formalism to study the combined effect of Hamiltonian evolution, local and collective decoherence, and measurement backaction in preparing nonclassical spin states for application in quantum metrology.
\end{abstract}

\maketitle

\section{Introduction}
Atom-light entanglement of spin ensembles is an important tool in a range of applications in quantum information processing (QIP), including quantum memory \cite{atomic_memory2018}, quantum metrology~\cite{Pezze2018}, quantum simulation~\cite{Schleier-Smith2020}, quantum computing~\cite{computing1}, and quantum feedback~\cite{Munoz2010}. In many of these applications the spins are uniformly coupled to a given spatial mode of an entangling probe-field, in which case quantum information is encoded in the collective spin of the permutationally invariant symmetric subspace. Examples include the creation of spin squeezed states~\cite{Leroux2010, Mitchell2012, kasevich2016, Cox2016} and Dicke states~\cite{Polzik2014, Vuletic2015exp} of the collective spin based on the backaction induced by measurement of the probe-field. The latter is an example of a nonGaussian state -- a particularly useful resource in QIP -- which can be created through, e.g., the measurement backaction on the atoms following the counting of single photons~\cite{vuletic_theory,vuletic_nature}.

All QIP applications are limited by decoherence. For the atom-light interface the major source of decoherence is due to the loss of photons that are entangled with the atoms but not measured. This loss could arise from finite detector efficiency, (and thus loss of forward scattered photons), or by diffuse scattering of photons out of the probe mode. The former leads to collective spin jumps, which are easily modeled as the system remains in the symmetric subspace. The latter is a local jump process associated with spin flips (i.e., optical pumping). In a given quantum trajectory of the system, the environment retains information about which atomic spin flipped, and thus the system is no longer permutationally symmetric. Exact modelling of the many-body system in the presence of this ``local decoherence'' becomes intractable, as the full Hilbert space grows exponentially with the number of atoms. A good approximate description of local decoherence is thus necessary to evaluate the usefulness of the atom-light interface, particularly as the nonGaussian states will be most sensitive to effects such as optical pumping.

An important step in this direction was made by Chase and Geremia who defined so-called ``collective states'' for $N$-body spin systems~\cite{chase}. These mixed states are invariant not only under unitary processes that are permutationally symmetric but also under dissipative processes such as those with local jump processes that are uniform across the ensemble. For weak decoherence this allows us to restrict the Hilbert space to subspaces that grow polynomially with the number of spins, a substantial reduction over the exponentially large total space. Zhang {\em et al.} used this collective states formalism to simulate the superradiant dynamics of ensembles of atoms in the presence of collective and individual atomic decay processes~\cite{zhang}. Their simulations were based on the Monte-Carlo wave function method \cite{molmer_mcwf}, using ensembles of quantum trajectories to simulate the Lindblad master equation, and made possible by the limited Hilbert space of the collective states for a moderate number of atoms.

In cases where large ensembles, e.g., $N\sim 10^6$ atoms, are used to enhance the cooperativity, even this reduction of the Hilbert space size due to collective states will often be insufficient to allow for tractable numerical simulations and additional approximations will be necessary. For the case of the collective spin in the symmetric subspace, the Holstein-Primakoff transformation~\cite{PhysRev.58.1098} -- where the spin degrees of freedom are replaced by those of single bosonic mode -- well approximate the quantum fluctuations around the mean field. The spin ensemble is then accurately modeled in a Hilbert space with a continuous variable representation over a limited phase space.

In this work we extend the phase space description of spin ensembles beyond the usual Holstein-Primakoff approximation on the symmetric subspace. This is done by applying the phase space description to the case of collective states, which allows for an open systems description of the atom-light interface in the presence of local, but uniform, decoherence. In turn, the description enables us to model the master equation in the Wigner-Weyl representation as a diffusion or Fokker-Planck equation, in some cases with analytic solutions. Our results for the open system generally allows us to study the degree of quantum advantage that we can expect in the creation of nonclassical states, e.g., through the reduction of the quantum Fisher information seen in smoothing of sub-Planck structure of the Wigner function.

The remainder of this article is organized as follows. In Sec.~\ref{sec:optical_pumping} we develop a formalism based on collective states to show how the state of an atomic spin ensemble undergoing optical pumping is approximated by the equation of motion on bosonic phase space. In Sec.~\ref{sec:operators} we further expand our formalism to include operator dynamics, and we discuss a choice of frame in which both operators and states are time dependent. In Sec.~\ref{sec:rig} we rigorously prove the results of Sec.~\ref{sec:optical_pumping} and go beyond optical pumping by generalizing the formalism to other types of local decoherence. In Sec.~\ref{sec:qfi} we demonstrate how our formalism can be used to quantify the metrological usefulness of a given state by deriving a correspondence between the quantum Fisher information of the atomic spin ensemble and that of the bosonic phase space description. Finally, we conclude on our work in Sec.~\ref{sec:conclusion}.

\section{Optical Pumping in the Collective State Basis}
\label{sec:optical_pumping}
We study the atom-light interface in which a probe laser mode is entangled with an atomic ensemble of spins through a dispersive interaction. For concreteness we consider the Faraday interaction whereby the rotation of the photon polarization is determined by the magnetization of the atoms along the propagation direction~\cite{TheLongPaper}. Assuming all $N$ atoms in the ensemble are uniformly coupled to the probe, the interaction entangles the Stokes vector of the polarization with some component $\hat J_\alpha = \sum_{i=1}^N  \hat\sigma_\alpha/2$ of the collective magnetization.  A measurement of the rotation of the light polarization on the Poincar\'{e} sphere thus provides a quantum nondemolition (QND) measurement of $\hat J_\alpha$ and the associated backaction.

\begin{figure}[t]
\includegraphics[width=.8\linewidth]{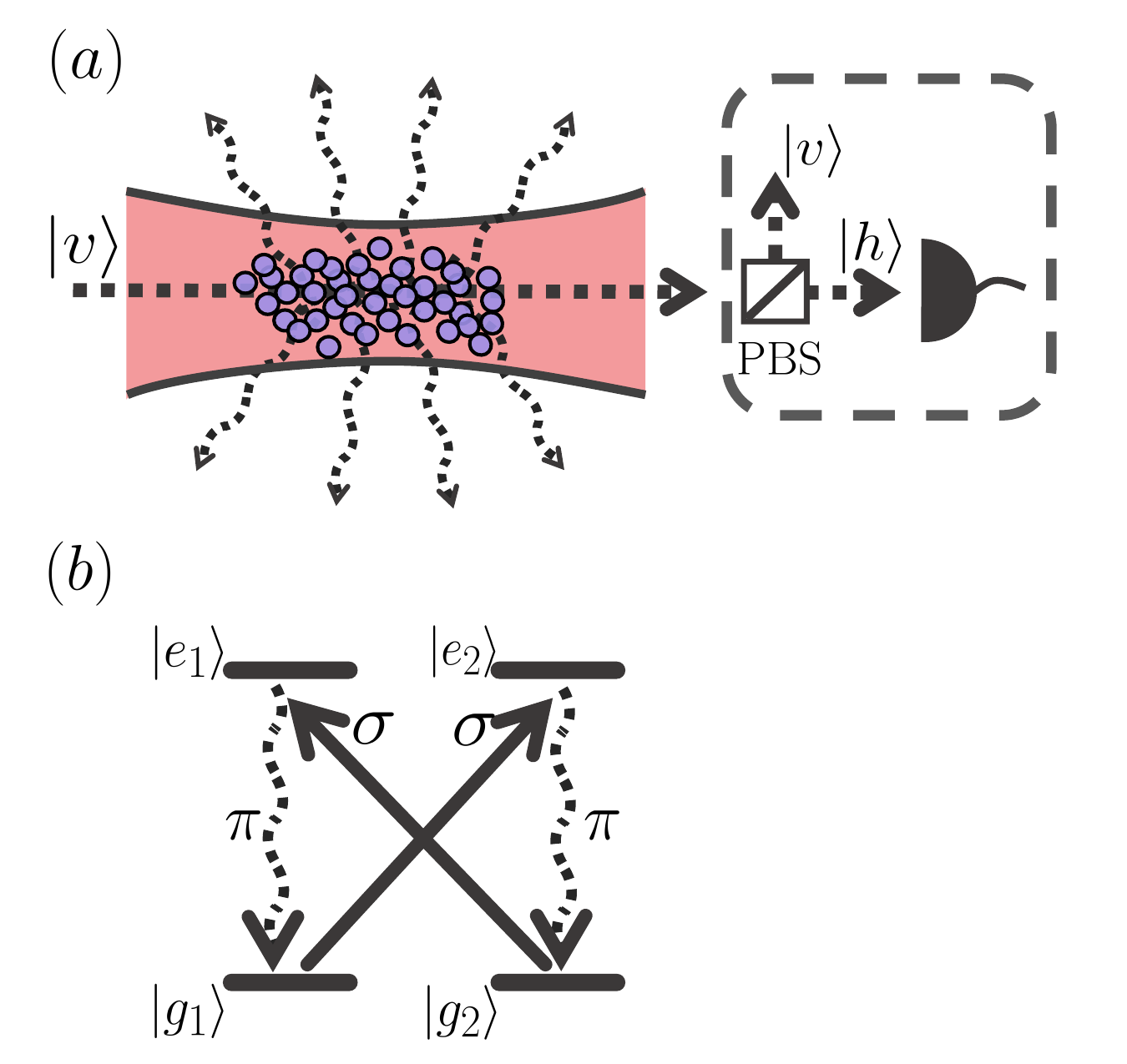}
\caption{(a)~Schematic of the atom-light interface.  A laser probe mode uniformly interacts with an ensemble of atoms. Wavy lines denote diffuse local scattering of photons out of the mode, causing decoherence. Due to a Faraday interaction, a single photon from the probe can be collectively scattered from $\ket{v}$-polarization to $\ket{h}$-polarization and detected, as shown in the dashed box.  The resulting measurement backaction will create a nonGaussian state. (b)~Schematic level diagram for optical pumping that causes spins spin flips, $\ket{\uparrow}=\ket{g_2} \leftrightarrow \ket{\downarrow}=\ket{g_1}$, for the spin-quantization axis perpendicular to $v$.}
\label{fig:main_figures}
\end{figure}

Figure~\ref{fig:main_figures} shows a simplified atomic model for linearly polarized light coupling dispersively on a $S_{1/2}\rightarrow P_{1/2}$ transition. The difference in the atom-light coupling for $\sigma_\pm$ polarization leads to a Faraday effect. Spontaneous emission of $\pi$-polarized light in turn causes optical pumping and spin-flips $\ket{\uparrow}\leftrightarrow\ket{\downarrow}$ with equal scattering rates $\gamma$. This is necessarily associated with diffuse scattering out of the probe mode. In the absence of any other Hamiltonian interaction, the master equation describing this optical pumping is given by,
\begin{align}
\party{}{t}\hat\rho &= \sum_{i=1}^N \mathcal{D}\Big[\sqrt{\gamma}\hat{\sigma}_+^{(i)}\Big]\hat{\rho} + \mathcal{D}\Big[\sqrt{\gamma}\hat{\sigma}_+^{(i)}\Big]\hat{\rho} \nonumber\\
&= -\gamma N \hat{\rho} + \gamma \sum_{i=1}^N \Big[ \hat{\sigma}_+^{(i)} \hat{\rho} \hat{\sigma}_-^{(i)} + \hat{\sigma}_-^{(i)} \hat{\rho} \hat{\sigma}_+^{(i)}\Big],\label{master1}
\end{align}
where $\gamma$ is the photon scattering rate, $N$ is the total number of atomic spins, and where we have defined the superoperator $\mathcal{D}[\hat c]\hat{\rho} = \hat{c} \hat{\rho} \hat{c}^\dagger - \frac{1}{2} \hat{c}^\dagger \hat{c} \hat\rho - \frac{1}{2}\hat{\rho} \hat{c}^\dagger \hat{c}$. The general solution to this equation is intractable, as local jump operators take the state out of the symmetric subspace, requiring one to keep track of the full Hilbert space of dimension $2^N$. Instead, we solve Eq.~(\ref{master1}) by making use of the formalism developed in~\cite{chase}. The Hilbert space is decomposed into the irreducible subspaces (irreps) of $\text{SU}(2)$, with eigenvectors $\ket{J,M,\alpha}$ which are the (generally degenerate) simultaneous eigenvectors of $\hat J^2$ and $\hat J_z$. The quantum number $\alpha$ labels the degenerate irreps such that $\braket{J,M,\alpha}{J,M,\alpha'}=0$ for $\alpha\neq\alpha'$. The degeneracy of the irreps with the same $J$ quantum number is~\cite{chase}
\begin{equation}
    d_N^J=\frac{N!(2J + 1)}{(N/2 - J )!(N/2 + J + 1)!}.
\end{equation}
The symmetric subspace, where $J=N/2$, is nondegenerate and of dimension $N+1$.

For a master equation with local jump operators applied at equal rates across all sites, one can write the density operator in terms of the \textit{collective state} basis \cite{chase,zhang}, whose elements
\begin{equation}
\col{J,M,N}{J,M',N}=\frac1{d_N^J}\sum_{\alpha}\ketbra{J,M,\alpha}{J,M^\prime,\alpha}\label{collective}
\end{equation}
are sums over the outer products of irrep eigenvectors $\ket{J,M,\alpha}\bra{J,M^\prime,\alpha}$, equally weighted over the degeneracy subspaces labeled by $\alpha$. The ability to use collective states to describe the effects of the light-matter interaction results from the fact that the total state remains invariant under permutation of the spins, which reduces the necessary Hilbert space significantly. We emphasize that the collective state $\col{J,M,N}{J,M',N}$ defined in Eq.~(\ref{collective}) is not an outer product of two pure states; we denote that by the use of double bars. However, the following relationship between the collective states and the sum of outer product
\begin{align}\label{coll_states}
    &\frac1{d_N^J}\sum_{\alpha,\alpha'}\ketbra{J,M,\alpha}{J,M',\alpha'} \nonumber\\
    &=\col{J,M,N}{J,M',N}+\frac1{d_N^J}\sum_{\alpha\neq\alpha'}\ketbra{J,M,\alpha}{J,M',\alpha'},
\end{align}
reveals that the collective states as defined in Eq.~(\ref{collective}) are the diagonal elements in $\alpha$ of the sum of outer products $\sum_{\alpha,\alpha'}\ketbra{J,M,\alpha}{J,M',\alpha'}$.

For expectation values of observable operators of the form $\hat C=\sum_{i=1}^N\hat c^{(i)}$, where $\hat c^{(i)}$ is a local Hermitian operator acting on the $i$th spin, there is no contribution from the off-diagonal terms $\alpha \neq \alpha^\prime$ in Eq.~(\ref{coll_states}) (see App.~\ref{sec:outer_product_proof} for details). Likewise, the off-diagonal terms do not contribute to expectation values of products of such operators. For the light-matter interaction considered here, the operators of interest -- which couple to the probe uniformly -- are of this particular form as they respect the permutation symmetry. Thus we need only concern ourselves with the diagonal (collective state) terms $\col{J,M,N}{J,M',N}$ in Eq.~(\ref{collective}). In the remainder of this work we therefore neglect the off-diagonal terms and treat the collective states $\col{J,M,N}{J,M',N}$ as being equal to the outer product of pure states $\sum_{\alpha,\alpha'}\ketbra{J,M,\alpha}{J,M',\alpha'}$. For simplicity we will drop the system size label $N$ from the collective states going forward, i.e., $\col{J,M}{J,M'}\equiv\col{J,M,N}{J,M',N}$.

\begin{figure}[]
    \centering
    \includegraphics[scale=0.32]{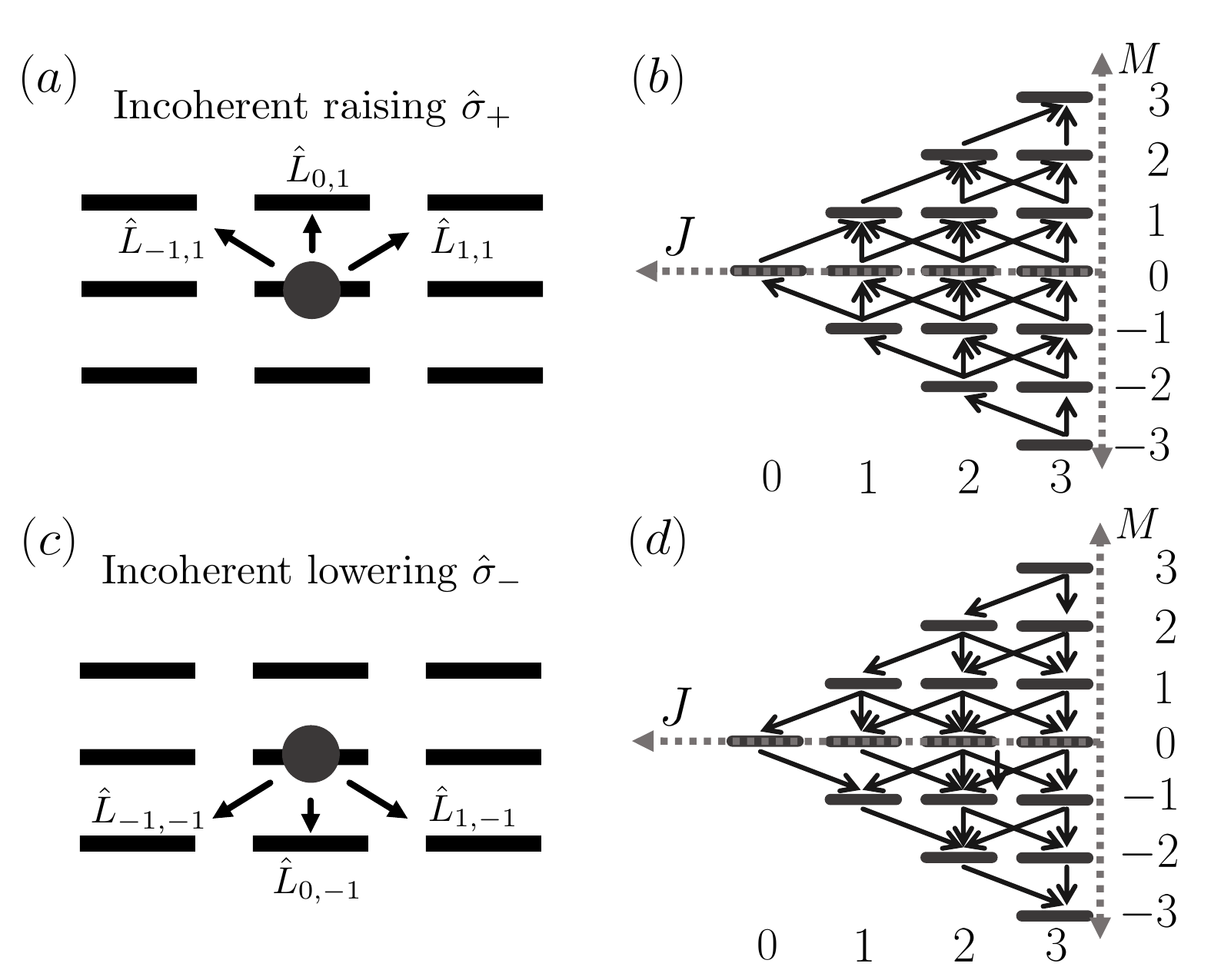}
    \caption{Population transfer between collective states during scattering events (local decoherence). (a)~Diagram showing how the raising operator moves population from a single collective state to the three neighboring collective states above it. This corresponds to three jump operators $\hat{L}_{j,1}$ with $j=-1,0,1$ the difference in the $J$ quantum number. (b)~An example for $N=6$ spins of population transfer between the collective states labeled $(J,M)$  undergoing raising events. (c)~Similarly, for the lowering operator, population is moved from one collective state to the three lower-lying neighbors, and we define three jump operators $\hat{L}_{j,-1}$ with $j=-1,0,1$, and the corresponding jumps in the ladder are shown in (d).}
    \label{fig:pumping}
\end{figure}

The Hilbert space can now be represented by these collective states, which can be visualized as a hierarchy of states labeled by $J$ and $M$ as illustrated in Fig.~\ref{fig:pumping}(b) and Fig.~\ref{fig:pumping}(d). This hierarchy is particularly well-suited to modeling optical pumping since emission events transfer population from $\col{J,M}{J,M}$ to neighboring states as depicted by the arrows in Fig.~\ref{fig:pumping}. Specifically, an optical pumping event can lower/raise the state within a given collective state subspace $J$ and/or simultaneously jump the state to a subspace with neighboring value of $J$. If we retain only the irreps with $J=N/2$ and $J=N/2-1$, the collective state representation reduces the dimension of the Hilbert space from exponential $d=2^N$ to polynomial $d\sim N^2$ in the system size $N$.

Chase and Geremia~\cite{chase} showed that one can write the action of a sum of local operators on a collective state as a sum over neighboring collective states. Explicitly, this may be written as
\begin{align}
    \sum_i&\hat\sigma_q^{(i)}\col{J,M}{J,M'}{\hat\sigma_r}^{(i)\dagger}\nonumber\\
    &\hspace{0.6cm}=A_{q,r}\col{J,M+q}{J,M'+r}\nonumber\\
    &\hspace{1.0cm}+B_{q,r}\col{J-1,M+q}{J-1,M'+r}\nonumber\\
    &\hspace{1.0cm}+D_{q,r}\col{J+1,M+q}{J+1,M'+r},
\end{align}
where the indices $q$ and $r$ take values $+1$, $-1$, and $0$ corresponding to the jump operators $\hat\sigma_+$, $\hat\sigma_-$, and $\hat\sigma_z$, respectively. $A$, $B$, and $D$ are real coefficients that depend on $N$, $J$, and $M$. Using this formalism one can rewrite Eq.~(\ref{master1}) by defining jump operators $\hat L_{j,q}$ which act on the density matrix $\hat{\rho}$ in the collective state basis. The subscript $j$ takes values $-1$, $0$, or $1$, and the subscript $q$ is either $-1$ or $1$. Altogether this yields six possible collective states to which population from $\col{J,M}{J,M'}$ can move under application of the local operators $\hat\sigma_+$ and $\hat\sigma_-$, as illustrated in Fig.~\ref{fig:pumping}. In the collective state basis the Lindblad master equation reads
\begin{equation}\label{master2}
\party{}{t}\hat\rho  =-\gamma N\hat\rho  +\gamma \sum_{j,q}\hat L_{j,q}\,\hat\rho  \, \hat L_{j,q}^\dagger,
\end{equation}
with the sum now running over only 6 terms instead of $N$. Equation~(\ref{master2}) will serve as the starting point for the remainder of our analysis.

\subsection{Solving for an initial spin coherent state}
Solving Eq.~(\ref{master2}) for large spin ensembles becomes numerically intractable, as $\hat\rho  $ is a matrix whose dimensions are proportional to $N^2\times N^2$. However, from Eq.~(\ref{master2}) we obtain the following rate equations for the population $p_{JM}$ in the $\col{J,M}{J,M}$ state, assuming that the initial state contains no coherence in the collective state basis:
\begin{align}
\frac d{dt}p_{J,M}(t)=&-\gamma Np_{J,M}(t)\nonumber\\
&+\sum_{j,q} \gamma \, g_{j,q}\;p_{J+j, M+q}(t). \label{ODE}
\end{align}
The six coefficients $g_{j,q}$ read,
\begin{subequations}
\begin{align}\label{g_first}
g_{1,1}&=\frac{(J+M+1)(J+M+2)(N+2J+4)}{4(2J+3)(J+1)},\\
g_{1,-1}&=\frac{(J-M+1)(J-M+2)(N+2J+4)}{4(2J+3)(J+1)},\\
g_{0,1}&=\frac{(J-M)(J+M+1)(N+2)}{4J(J+1)},\\
g_{0,-1}&=\frac{(J+M)(J-M+1)(N+2)}{4J(J+1)},\\
g_{-1,1}&=\frac{(J-M)(J-M-1)(N-2J+2)}{4J(2J-1)},\\
\label{g_last}g_{-1,-1}&=\frac{(J+M)(J+M-1)(N-2J+2)}{4J(2J-1)}.
\end{align}
\end{subequations}
We can find a simple closed form solution to Eq.~(\ref{ODE}) when the initial state is a spin coherent state along the $z$-axis, $\hat\rho (t=0)=(\ketbra{\uparrow}{\uparrow})^{\otimes N} = \ket{J,J}\bra{J,J}$. For this particular state the initial populations are $p_{J,M}(0)=\delta_{J,N/2}\,\delta_{M,N/2}$, and we find that the density operator at a later time~$t$ is
\begin{equation}\label{pjms_master}
\hat\rho (t)=\sum_{J,M}p_{JM}(t)\,\col{J,M}{J,M},
\end{equation}
where the time-dependent population $p_{JM}(t)$ take the form
\begin{align}\label{pjms}
    p_{JM}(t)&=d_J^N \left(\frac{1+e^{-2\gamma t}}{2}\right)^{\frac{N}{2}+M} \left(\frac{1-e^{-2\gamma t}}{2}\right)^{\frac{N}{2}-M}.
\end{align}
Equation~(\ref{pjms}) follows by considering the total population in states labeled by the quantum number $M$, and then restricting the population to only the states labeled by the set of quantum numbers ($J$, $M$). Consider first a single qubit in the initial state $\ket{\uparrow}$. After some time $t$ the probabilities of being in the $\ket\uparrow$-state and the $\ket\downarrow$-state are given by
\begin{align}
    P_\uparrow=\frac{1 + e^{-2\gamma t}}{2} \;\; \text{and} \;\; P_\downarrow=\frac{1 - e^{-2\gamma t}}{2},
\end{align}
respectively. Therefore, the probability for a system of $N$ spin-1/2 particles to be in state labeled by $M$ after some time $t$ is
\begin{equation}
    p_M(t)=\binom{N}{\frac N2-M}P_\uparrow^{\frac N2+M}P_\downarrow^{\frac N2-M},
\end{equation}
where $\frac N2+M$ is the number of spins in $\ket\uparrow$ and $\frac N2-M$ is the number of spins in $\ket\downarrow$. The binomial coefficient counts the number of configurations of $N$ particles such that the total spin along $J_z$ sums to $M$.

Next we wish to consider only the population in states which can be labeled by the pair ($J$, $M$). This population is obtained by multiplying the total population $p_M(t)$ by the ratio of states labeled by $J$ and $M$ to the total number of states labeled by $M$ alone. This ($J$, $M$) population is
\begin{equation}
    p_{J,M}(t)=\frac{d^N_J}{\sum_{J'=M}^N d^N_{J'}}p_M(t)=d^N_J P_\uparrow^{\frac N2+M}P_\downarrow^{\frac N2-M},
\end{equation}
identical to the form presented in Eq.~(\ref{pjms}). In multiplying by the ratio $d^N_J/\sum_{J'=M}^N d^N_{J'}$ we have assumed that the population in $\ketbra{J,M,\alpha}$ is equal to the population in $\ketbra{J',M,\alpha'}$ for all $J$, $J^\prime$, $\alpha$, and $\alpha^\prime$. While populations with $J=J^\prime$ and $\alpha\neq\alpha^\prime$ must be equal due to permutation symmetry, it is not obvious that populations associated with $J\neq J'$ would also be equal. Nonetheless, this leads to a solution which satisfies Eq.~(\ref{ODE}) for the initial condition of a spin coherent state.

With the results in Eqs.~(\ref{pjms_master}-\ref{pjms}) established we make the Holstein-Primakoff approximation~\cite{PhysRev.58.1098}, which is valid for large $N$ when $\langle\hat J_z\rangle\sim J\sim N/2$. The Holstein-Primakoff approximation allows us to represent the collective spin algebra using the algebra of a bosonic mode, mapping Dicke states to Fock states $\ket{J,M}\to\ket{n=J-M}$ and mapping operators $\hat J_x$ and $\hat J_y$ as
\begin{equation}
\frac{\hat J_x}{\sqrt J}\to \hat X\;\;\;\;\text{and}\;\;\;\;\frac{\hat J_y}{\sqrt J}\to \hat P,
\end{equation}
such that the canonical commutation relation is preserved, i.e. $[\hat X,\hat P]\to\left[\frac{\hat J_x}{\sqrt J},\frac{\hat J_y}{\sqrt J}\right]=i\frac{\hat J_z}{J}\approx i$. We generalize the mapping of Dicke states to Fock states by allowing collective states to be mapped to Fock states as $\col{J,M}{J,M'}\to\ketbra{n=J-M}{n'=J-M'}$. We recall that this is a valid mapping due to the collective states behaving as the outer product of pure states. This particular mapping of collective states to Fock states is straightforward when considering cases where no coherence exists between irreps. Applying this mapping our density operator $\hat\rho $ becomes 
\begin{equation}
\hat\rho _{\text{HP}}(t)=\sum_n p_n(t)\ketbra n,
\end{equation}
where by using Eq.~(\ref{pjms}) the population $p_n(t)$ can be written as
\begin{align}
p_n(t)&=\lim_{N\to\infty}\sum_Jp_{J,J-n}(t) \nonumber\\
&=\tanh^n(\gamma t) [1-\tanh(\gamma t)]. \label{eq:NlimitHP}
\end{align}
A detailed derivation of Eq.~(\ref{eq:NlimitHP}) can be found in App.~\ref{sec:p_n}.

The distribution described by $p_n(t)$ is that of a thermal state
\begin{equation}
\hat\rho _{\text{HP}}(t)=\sum_{n=0}^\infty \frac{\langle n(t) \rangle^n}{[1+\langle n(t)\rangle]^{n+1}} \ket{n}\bra{n},
\end{equation}
with time-dependent mean photon number
\begin{equation}
    \langle n(t) \rangle = \frac{\tanh(\gamma t)}{1-\tanh(\gamma t)} = \frac{1}{2}\left( e^{2\gamma t} -1\right).
\end{equation}
Using this result and the Holstein-Primakoff approximation, the master equation in the Wigner-Weyl representation takes the form of a diffusion equation with a time-dependent diffusion coefficient $D(t) = \gamma [\langle n(t) \rangle + \frac{1}{2}]$, which may be written on the form
\begin{align}
\party{}{t}W_{\text{TH}}(X,P,t)=& D(t)\, \nabla^2 W_\text{TH}(X,P,t) \nonumber\\
=&\frac{\gamma}{2} e^{2\gamma t} \left(\party{^2}{X^2}W_{\text{TH}}+\party{^2}{P^2}W_{\text{TH}}\right). \label{W_th}
\end{align}
Note that since this was derived for an initial spin coherent state, Eq.~(\ref{W_th}) is seemingly only valid when $\expval{n}=0$ at $t=0$. However, we will show in Sec.~\ref{sec:rig} that Eq.~(\ref{W_th}) holds for arbitrary initial states, including cases where $\expval{n}\neq 0$ at the initial time~$t=0$.

The mean excitation number $\expv{n}$ of Wigner function $W_\text{TH}$ increases with time~$t$, which corresponds to a decrease in $M$ since the Holstein-Primakoff mapping is $\ket{J,M}\to\ket{n=J-M}$. Additionally, the Holstein-Primakoff approximation treats all states with the same $J-M$ value identically, even if they have different values of $J$. As a consequence the decreasing value of $J$ does not manifest in the state. We shall see in Sec.~\ref{sec:operators} that the decreasing value of $J$  appears as a time dependence in the operators.

Although $W_{\text{TH}}$ can be used to study the evolution of an initial spin coherent state undergoing optical pumping, the variance of the quadratures of $W_{\text{TH}}$ increase with time, whereas the variance of the spin projections $\hat J_x$ and $\hat J_y$ are invariant with time~\footnote{For an initial spin coherent state, all spin-$1/2$ particles remain uncorrelated under incoherent optical pumping, hence the variance in $\hat J_x$ and $\hat J_y$ is the sum of the variances of all spins, which is constant in time.}. One may wish to obtain a description of the spin system in a bosonic mode in which $\hat X$ and $\hat P$ remain proportional to  $\hat J_x$ and $\hat J_y$ at all times. To this end we note that the Holstein-Primakoff approximation contains a proportionality constant $J$, the total spin polarization, which decreases with time as the mean spin decays due to optical pumping: $\expval{J}(t)=(N/2)e^{-2\gamma t}$. For sufficiently large $N$ we can make the approximation $J\approx\expval{J}(t)$, i.e., we replace $J$ in the Holstein-Primakoff approximation by its mean value (see App.~\ref{sec:p_n} for additional details). This allows us to make a nonunitary frame transformation and define a {\em generalized} Holstein-Primakoff approximation,
\begin{equation}\label{timeHP1}
\hat X\approx\frac{\hat J_x}{\sqrt{J(t)}}=\frac{\hat J_x}{\sqrt{N/2}}e^{\gamma t}\;\;\to\;\;\tildehat{X}\approx\frac{\hat J_x}{\sqrt{N/2}},
\end{equation}
such that $\tildehat X=\hat Xe^{-\gamma t}$, and similarly for $\hat P\to\tildehat P$. Note that under this frame transformation the commutation relations are not preserved, and thus the transformed quadratures should only be understood in the context of calculating expectation values.

By considering the bosonic mode in terms of the quadratures $\tildehat X$ and $\tildehat P$, we obtain a description in which the quadratures of the mode are directly proportional to the fluctuations of the spin projections at all times with proportionality factor $\sqrt{N/2}$. To find the equation of motion of the bosonic mode in the new frame, we perform a change of variables to $\widetilde X=Xe^{-\gamma t}$, $\widetilde P=Pe^{-\gamma t}$ in Eq.~(\ref{W_th}). We also make the transformation $e^{2\gamma t}\,W(\widetilde X e^{\gamma t},\widetilde P e^{\gamma t},t)\equiv \widetilde{W}(\widetilde X,\widetilde P,t)$ in order to preserve normalization. After the change of variables Eq.~(\ref{W_th}) becomes
\begin{equation}\label{fp}
\party{\widetilde{W}}{t}=\gamma\left(\party{}{\widetilde  X}\widetilde  X+\party{}{\widetilde P}\widetilde P\right)\widetilde{W}+\frac\gamma2 \left(\party{^2}{\widetilde  X^2}+\party{^2}{\widetilde P^2}\right)\widetilde{W},
\end{equation}
which takes the form of a Fokker-Planck equation with time-independent coefficients. This master equation is well-known and describes the evolution of a damped harmonic oscillator coupled to a zero temperature bath with boson loss rate $\gamma$.

It follows from Eq.~(\ref{fp}) that fluctuations in $\hat J_x$ and $\hat J_y$ evolve under a Fokker-Planck equation for an initial spin coherent state $\ket{\uparrow}^{\otimes N}$ along $\hat J_z$. From here on we will refer to the use of Eq.~(\ref{fp}) to represent optical pumping in the decaying frame as the ``Fokker-Planck picture''. Similarly, we will refer to the use of Eq.~(\ref{W_th}) to represent optical pumping in the fixed frame as the ``thermalizing picture''. Each picture comes with a different set of operators, and is represented in phase space by a different differential equation. Figure~(\ref{fig:comparison}) provides a comparison of the two pictures.

In the Fokker-Planck picture the mean excitation number $\expv{n}$ decreases, in contrast to the thermalizing picture in which $\expv{n}$ increases. Consequently, in the Fokker-Planck picture the effect of the decaying $J_z$ is no longer included in the Wigner function. The decreasing $M$-value instead appears in the evolution of the operators alongside the effect of the decreasing $J$-value, as we will explore further in Sec.~\ref{sec:operators}.

The frame choice depends on the particular quantity that we want to evaluate, with each picture providing some utility over the other. If one wishes to calculate the probability distributions of $\hat J_x$ and $\hat J_y$ under optical pumping, then the Fokker-Planck picture (Eq.~(\ref{fp}) and corresponding operators) is preferable since the bosonic marginals are directly proportional to the probability distributions of $\hat J_x$ and $\hat J_y$. However, we shall see in Sec.~\ref{sec:qfi} that if one instead wishes to calculate a quantity like the quantum Fisher information (QFI) then the thermalizing picture (Eq.~(\ref{W_th}) and corresponding operators) is preferable. This is because the Fock space structure in the thermalizing picture is approximately proportional to the structure of Dicke states in spin space. Finally, while Eq.~(\ref{fp}) was derived for an initial spin coherent state, we will show in Sec.~\ref{sec:rig} how to generalize this result to all initial states, and even to other forms of local decoherence.

\section{Bosonic operator transformations in two frames}\label{sec:operators}
\begin{figure*}
    \centering
    \includegraphics[scale=0.52]{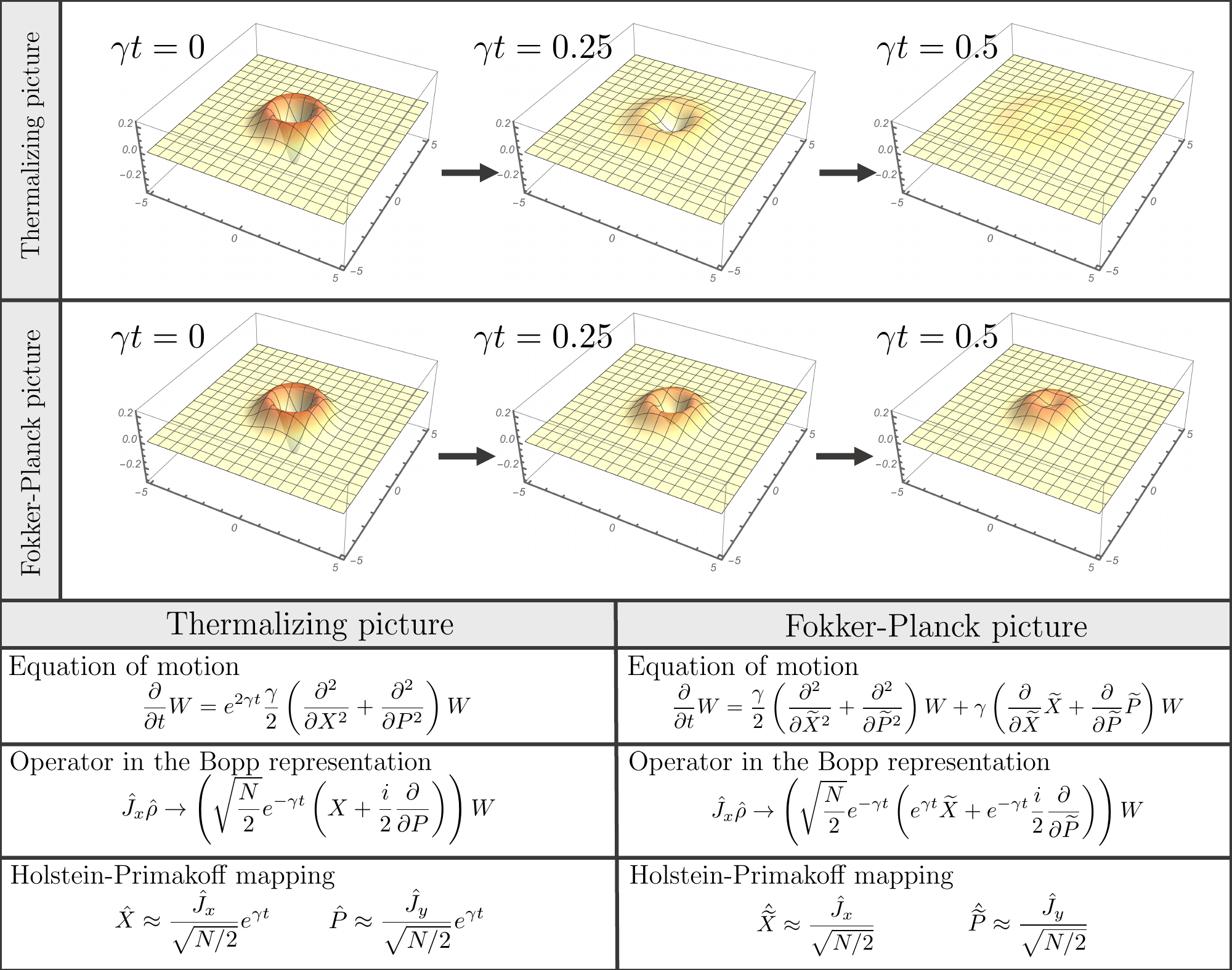}
    \caption{Comparison of two pictures (frames) to describe local decoherence as the evolution of a bosonic mode. Top panels: Time evolution in both the thermalizing (TH) picture and the Fokker-Planck (FP) picture for an initial first excited Fock state $\hat\rho =\ketbra1$, the bosonic approximation to the first Dicke state. In the thermalizing picture, the Wigner function diffuses over phase space, whereas in the Fokker-Planck picture it remains localized near the origin. Bottom panels: comparisons of the formalisms for each picture, including the equations of motion for the Wigner function, examples of differential operators, and the Holstein-Primakoff mappings between phase space operators and collective spin operators.}
    \label{fig:comparison}
\end{figure*}

As discussed in Sec.~\ref{sec:optical_pumping}, the dynamics of the atomic ensemble undergoing local optical pumping can be described in two frames or ``pictures.''  These pictures lead to different equations of motion for the system's Wigner function, and likewise different transformations of the operators in the Wigner-Weyl representation. In this section we discuss how to apply the frame transformations of Sec.~\ref{sec:optical_pumping} to the relevant operator representations, including both the differential operators associated with dynamical evolution and the Kraus operators associated with measurement backaction. As we shall see, the choice of picture has implications for the modeling of open system dynamics and measurement. Our findings from this section will be summarized in Fig.~(\ref{fig:comparison}).

A spin angular momentum operator acting on the space can be written in the generalized Holstein-Primakoff approximation using the time-dependent mapping in Eq.~(\ref{timeHP1}). As an example, $\hat J_x$ is transformed as 
\begin{equation}
\hat J_x\to\sqrt{\frac N2}e^{-\gamma t}\hat X.
\end{equation}
In the Wigner-Weyl representation, the action of an operator on the state is represented by a differential operator.  For example, using the \textit{Bopp representation}~\cite{BoppRep2010} the $\hat J_x$ operator in the thermalizing picture takes the form
\begin{align}
\label{therm_bopp}
\sqrt{\frac N2}e^{-\gamma t}\hat X \hat\rho  &\to\underbrace{\sqrt{\frac N2}e^{-\gamma t}\left(X+\frac i2\party{}{P}\right)}_{=:\mathcal{X}_{\text{TH}}} W(X,P),
\end{align}
where we have defined $\mathcal X_{TH}$ as a shorthand for the action of $\hat J_x$ on the Wigner function in the thermalizing picture. Similarly, in the Fokker-Planck picture we take Eq.~(\ref{therm_bopp}) and apply the same transformation of variables as in Eq.~(\ref{fp}) to find the operator form
\begin{align}
\label{fp_bopp}
\sqrt{\frac N2}e^{-\gamma t}X \hat\rho  &\to \underbrace{\sqrt{\frac N2}\left(\widetilde X+e^{-2\gamma t}\frac i2\party{}{\widetilde P}\right)}_{=: \mathcal X_{\text{FP}}} W(\widetilde X, \widetilde P).
\end{align}
Here we have defined $\mathcal X_{\text{FP}}$ as a shorthand for the action of $\hat J_x$ on the Wigner function, but now in the Fokker-Planck picture. Notice that in this picture the differential term decays exponentially, while $\widetilde X$ does not. This is a result of both the decaying $M$-value and the decaying $J$-value being contained in the operators in this picture.

Next we consider two examples that demonstrate the effects of open quantum systems in both pictures. We consider first the case in which both local optical pumping and {\em collective} decoherence are present. Collective decoherence occurs when the ensemble collectively scatters photons into the forward mode which are not measured due to, e.g., finite detector efficiency. This additional collective decoherence may be described in our Lindblad master equation Eq.~(\ref{master1}) by adding an additional term $\mathcal{D}[\sqrt{\kappa} \hat J_x]= \kappa \hat J_x \hat \rho \hat J_x -\frac{\kappa}{2} \{\hat J_x^2, \hat \rho \}$ with the jump operator $\sqrt\kappa \hat J_x$, where $\kappa$ is the rate of emission per atom into the mode that couples the collective spin operator $\hat J_x$ to the probe beam. In the thermalizing picture, the master equation corresponding to simultaneous local and collective decoherence may be written as
\begin{align}
    \party{}{t}W=&\mathcal X_{\text{TH}}^*\mathcal X_{\text{TH}}W -\frac{1}{2}\mathcal X_{\text{TH}}^2 W-\frac{1}{2}\mathcal X_{\text{TH}}^{*2}W+\text{TH}[W] \nonumber\\
    =&e^{-2\gamma t}\frac{\kappa N}{2}\party{^2}{P^2}W+\text{TH}[W],\label{coll_dec_therm}
\end{align}
where $\text{TH}[\;\cdot\;]$ is the mapping on the RHS of Eq.~(\ref{W_th}), representing local decoherence. Similarly in the Fokker-Planck picture one finds 
\begin{align}
    \party{}{t}W=&\mathcal X_{\text{FP}}^*\mathcal X_{\text{FP}}W
    -\frac{1}{2}\mathcal X_{\text{FP}}^2 W-\frac{1}{2}\mathcal X_{\text{FP}}^{*2}W+\text{FP}[W]\nonumber\\
    =&e^{-4\gamma t}\frac{\kappa N}{2}\party{^2}{\widetilde P^2}W+\text{FP}[W], \label{coll_dec_fp}
\end{align}
where $\text{FP}[\;\cdot\;]$ is the differential map on the RHS of Eq.~(\ref{fp}), again representing local decoherence. Note that in both Eq.~(\ref{coll_dec_therm}) and Eq.~(\ref{coll_dec_fp}) the first term on the RHS -- corresponding to collective decoherence -- is decaying. This is a consequence of optical pumping driving the state to the maximally mixed state.

The second example that we will consider here is that of a QND measurement of $\hat J_x$. Figure~\ref{fig:main_figures}(a) illustrates such a measurement setup where the atomic ensemble scatters photons into both the environment (leading to optical pumping) as well as into the forward mode parallel with the incident probe beam. Due to the Farday effect, the signal photons scattered into the forward mode have horizontal polarization $\ket{h}$ orthogonal to the vertical polarization $\ket{v}$ of the incident probe mode, as shown in Fig.~\ref{fig:main_figures}. The horizontally polarized probe photons are separated from the probe photons via a polarized beam splitter (PBS) and subsequently measured by a single photon detector. For simplicity we assume here a perfect PBS and unit detector efficiency.

The interaction unitary between the atoms and light is given by
\begin{equation}
    \hat U_\text{int}=e^{-i\sqrt{\kappa t/2}\hat P_L\otimes \hat J_x},
\end{equation}
where $\kappa$ is the rate of scattering into the detection mode per atom, $t$ is the duration of the single photon detection, and $\hat P_L$ is the quadrature operator on the detection $h$-mode of the light. From this interaction one can calculate the Kraus operator $\hat K_1$ associated with measuring a single photon 
\begin{equation}
    \hat K_1=\mel{1_L}{\hat U_\text{int}}{0_L}=e^{-\frac{\kappa t}{8}\hat J_x^2}\sqrt{\frac{\kappa t}{4}}\hat J_x.
\end{equation}
Converting this Kraus operator into a differential operator $\mathcal K$ in the Fokker-Planck picture we obtain
\begin{align}
    \mathcal K=&\exp[-\frac{\kappa Nt}{16}\left(\widetilde X+e^{-2\gamma t}\frac i2\frac{\partial}{\partial \widetilde P}\right)^2]\nonumber\\
    &\;\;\times \sqrt{\frac{\kappa N t}{8}}\left(\widetilde X+e^{-2\gamma t}\frac i2\frac{\partial}{\partial \widetilde P}\right).
\end{align}
The post measurement state $W_\text{post}(\widetilde X,\widetilde P,t)$ is therefore
\begin{equation}
    W_\text{post}(\widetilde X,\widetilde P,t)=\frac{\mathcal K^*\mathcal K \;W_{\text{vac}}(\widetilde X,\widetilde P)}{\int\mathrm d\widetilde X\mathrm d\widetilde P\; \mathcal K^*\mathcal K \;W_{\text{vac}}(\widetilde X,\widetilde P)},
\end{equation}
where $W_{\text{vac}}(\widetilde X,\widetilde P)$ is the Wigner function of the vacuum state. Note that in this example, the state does not evolve in time, a result of the fact that the vacuum is the steady state in the Fokker-Planck picture. The time dependence of the post measurement state in this example is therefore contained solely within the Kraus operator.

\section{Extension to arbitrary initial states}
\label{sec:rig}
In previous sections we assumed for simplicity that the initial state was a spin coherent state, and using this assumption we derived the analytical approximation of the system undergoing local optical pumping. In this section we prove a generalization of these results, namely that the evolution of the Wigner function for any initial state under local optical pumping is described by a Fokker-Planck equation. Additionally we show that there exists a family of local decoherence models which can be described by a Fokker-Planck equation. The results we will derive below thus allow us to extend our results of previous sections to arbitrary initial states.

To show this, we will first prove that any moment of an arbitrary linear combination of the collective spin operators $\hat J_x$ and $\hat J_y$ obeys a Fokker-Planck equation for all initial states. From this we will argue that the time evolution of the Wigner function for any initial state undergoing local optical pumping also obeys a Fokker-Planck equation. Our proof will require us to utilize generalized spherical tensor operators, which we will briefly review for completeness. These generalized spherical tensors then allow us to rewrite the time evolution of an arbitrary moment of collective spin operators on a more convenient form.

The generalized spherical tensor operators are defined as~\cite{gen_sphere_tensor}
\begin{align}
\hat{T}_{k,q}^{J',J}=&\sqrt{\frac{2k+1}{2J'+1}}\sum_{M,M'}C_{k,q,J,M}^{J',M'}\ketbra{J',M'}{J,M}, \nonumber\\
=&\sqrt{\frac{2k+1}{2J'+1}}\sum_{M,M'}C_{k,q,J,M}^{J',M'}\col{J',M'}{J,M},
\end{align}
where $C_{k,q,J,M}^{J^\prime,M^\prime} = \langle J^\prime, M^\prime \vert k, q, J, M \rangle$ are the Clebsch-Gordan coefficients and in the second line we have rewritten the spherical tensors in terms of the collective states used throughout this work. The generalized spherical tensor operators couple subspaces of different $J$-values, in contrast to the usual spherical tensor operators which couple only states with identical $J$-values. Importantly, the transformation property of the usual spherical tensors also holds for the generalized spherical tensor operators, namely that under rotation,
\begin{equation}
\hat{R}_\nu(\theta) \, \hat{T}_{k,q}^{J,J'}\, \hat{R}_\nu^\dagger (\theta)= \sum_{q'} \hat{T}_{k,q'}^{J',J} D_{q',q}^{k,\nu}(\theta), \label{eq:GenSphericalRotProperty}
\end{equation}
where $\hat{R}_\nu(\theta)=e^{-i\theta\hat{J}_\nu}$ is the SU(2) rotation operator around the $\nu$-axis by the angle $\theta$, and $D^k$ is the corresponding Wigner $D$-matrix~\cite{Sakurai2011}.

The six jump operators derived previously in Sec.~\ref{sec:optical_pumping} for optical pumping can be expanded in these generalized spherical tensors as
\begin{equation}
\hat L_{j,q}=\sum_J\Lambda_{j,q}^{J,N}\;\hat{T}_{1,q}^{J+j,J}, \label{eq:JumpExpansionGenSpherical}
\end{equation}
where $j$ and $q$ are the changes in the $J$ and $M$ quantum numbers. The coefficients $\Lambda_{j,\;q}^{J,N}$ are
\begin{subequations}
\begin{align}
\Lambda_{1,q}^{J,N}&=-q\sqrt{\frac{(N-2J)(2J+3)}{6}},\\
\Lambda_{0,q}^{J,N}&=q\sqrt{\frac{(N+2)(2J+1)}{6}},\\
\Lambda_{-1,q}^{J,N}&=q\sqrt{\frac{(2J+N+2)(2J-1)}{6}}.
\end{align}
\end{subequations}
Consider now the time evolution of the operator $\hat J_x^n$ for an arbitrary integer $n \geq 1$, given by the adjoint master equation
\begin{align}
\frac{\mathrm{d}}{\mathrm{d}t} \hat J_x^n(t) =& \sum_{j,q}\hat L_{j,q}^\dagger \hat J_x^n\hat L_{j,q} \nonumber\\
&- \frac{1}{2}\left\{\hat L_{j,q}^\dagger \hat L_{j,q}, \hat J_x^n\right\} \nonumber\\
=&\sum_{j,q}\hat L_{j,q}^\dagger \hat J_x^n\hat L_{j,q}-\frac N2\hat J_x^n, \label{eq:simplifiedAME}
\end{align}
where in the second equality we have simplified the anti-commutator part of the master equation. It follows that in order to solve Eq.~(\ref{eq:simplifiedAME}) we need to simplify the first term $\sum_{j,q}\hat L_{j,q}^\dagger \hat J_x^n\hat L_{j,q}$ further.

To this end we apply unitary rotation $\hat{R}_y(\frac{\pi}{2})=\exp (-i\pi \hat{J}_y/2)$ to align $\hat J_x^n$ with the $z$-axis, since $\hat J_z^n$ is diagonal in the collective state basis. Using first the expansion Eq.~(\ref{eq:JumpExpansionGenSpherical}) and then the rotation property Eq.~(\ref{eq:GenSphericalRotProperty}) we find that
\begin{widetext}
\begin{align}
\nonumber \hat{R}_y\left(\frac{\pi}{2}\right) \Bigg(\sum_{j,q}\hat L_{j,q}^\dagger \hat J_x^n\hat L_{j,q}\Bigg)\hat{R}_y^\dagger\left(\frac{\pi}{2}\right) 
&=\hat{R}_y\left(\frac{\pi}{2}\right)\left(\sum_{J,j,q}\left(\Lambda_{j,q}^{J,N}\right)^2\left(\hat{T}_{1,q}^{J+j,J}\right)^\dagger \hat J_x^n\; \hat{T}_{1,q}^{J+j,J}\right)\hat{R}_y^\dagger\left(\frac{\pi}{2}\right) \nonumber\\
&=\sum_{J,j,q,\lambda,\lambda'}\!\!\left(\Lambda_{j,q}^{J,N}\right)^2\;D_{\lambda,q}^{1,y}\left(\frac{\pi}{2}\right)\;D_{\lambda',q}^{1,y}\left(\frac{\pi}{2}\right)\left(\hat{T}_{1,\lambda}^{J+j,J}\right)^\dagger\, \hat J_z^n\, \hat{T}_{1,\lambda'}^{J+j,J}. \label{rotated}
\end{align}
The resulting expression in Eq.~(\ref{rotated}) is diagonal in the collective state basis. This can be seen by noting that in the rotated frame, the jump operators for the $i^{th}$ site are $\hat\sigma_z^{(i)}$ and $\hat\sigma_y^{(i)}$. Acting on $J_z^n$ with these rotated jump operators from both sides leaves it diagonal in the collective state basis. The diagonal elements of Eq.~(\ref{rotated}) may therefore be written as 
\begin{align}
    \nonumber\bra{J,M}\hat{R}_y\left(\frac{\pi}{2}\right)&\left(\sum_{j,q}\hat L_{j,q}^\dagger \hat J_x^n\hat L_{j,q}\right)\hat{R}_y^\dagger\left(\frac{\pi}{2}\right)\ket{J,M}\\
    =&\sum_{j,q}\left(\frac{3}{2J+2j+1}\right)\left(\Lambda_{j,q}^{J,N}\right)^2\bigg[\left(C_{1,1,J,M}^{J+j,M+1}D^1_{1,\Delta M}(\pi/2,0)\right)^2(M+1)^n\nonumber\\
    &+\left(C_{1,0,J,M}^{J+j,M}D^1_{0,\Delta M}(\pi/2,0)\right)^2M^n+\left(C_{1,-1,J,M}^{J+j,M-1}D^1_{-1,\Delta M}(\pi/2,0)\right)^2(M-1)^n\bigg]\nonumber\\
    =&\frac14(N-2M)(M+1)^n+2NM^n+(N+2M)(M-1)^n. \label{eq:JxRotatedJumpedSimplification}
\end{align}
To obtain the final equality of Eq.~(\ref{eq:JxRotatedJumpedSimplification}) we evaluated the sums over $j$ and $q$ by inserting the explicit forms of the Clebsch-Gordan coefficients $C$ and the Wigner $D$-matrix. Using the diagonal matrix elements in Eq.~(\ref{eq:JxRotatedJumpedSimplification}), we can express Eq.~(\ref{rotated}) in terms of collective states as
\begin{align}
    \hat{R}_y\left(\frac{\pi}{2}\right)\Bigg(\sum_{j,q}&\hat L_{j,q}^\dagger \hat J_x^n\hat L_{j,q}\Bigg)\hat{R}_y^\dagger\left(\frac{\pi}{2}\right) \nonumber\\
    =&\frac14\sum_{J,M}\Big[(M+1)^n(N-2M)+(M-1)^n(N+2M)+2NM^n\Big]\col{J,M}{J,M},
\end{align}
This expression may be further simplified by using the binomial theorem and applying the inverse rotation $\hat{R}_y^\dagger\left(\frac{\pi}{2}\right)$ to recover $\hat J_x^n$. We find that
\begin{align}
\sum_{j,q}\hat L_{j,q}^\dagger \hat J_x^n\hat L_{j,q}=&\sum_{J,M}\Bigg(\frac{N}{2}M^n+\sum_\mu \binom{n}{\mu} \bigg[\frac{N}{4} M^\mu(1+(-1)^{n+\mu})+\frac{1}{2} M^{\mu+1}((-1)^{n+\mu}-1)\bigg]\Bigg)\nonumber\\
&\times\hat{R}_y^\dagger\left(\frac{\pi}{2}\right)\col{J,M}{J,M} \hat{R}_y\left(\frac{\pi}{2}\right).
\end{align}
\end{widetext}
Finally, since $\hat{R}_y^\dagger\left(\frac{\pi}{2}\right) \sum_M M^n \col{J,M}{J,M} \hat{R}_y\left(\frac{\pi}{2}\right) = \hat J_x^n$ we obtain the result
\begin{equation}
\label{jx_expression}\sum_{j,q}\hat L_{j,q}^\dagger\, \hat J_x^n \, \hat L_{j,q}=\sum_{p=0}^{n}c_p \,\hat J_x^p,
\end{equation}
where
\begin{align}
c_p=&\frac{1}{4} \left[(-1)^{n+p}+1\right] \left[N \binom{n}{p}-2 \binom{n}{p-1}\right] \nonumber\\
&+\delta_{p,n}\frac N2.
\end{align}
Equation~(\ref{jx_expression}) reveals that the sum over jump terms in the adjoint master equation Eq.~(\ref{eq:simplifiedAME}) is a linear combinations of only $\hat{J}_x^p$ terms with $p\leq n$.

Substituting Eq.~(\ref{jx_expression}) into the full adjoint master equation Eq.~(\ref{eq:simplifiedAME}) for $\hat J_x^n$, we find for $n$ even that $\langle \hat{J}_x^n \rangle$ evolves in time according to
\begin{align}
\frac{\mathrm{d}}{\mathrm{d}t}\expval{\hat J_x^n}=&-n\gamma\expval{\hat J_x^n}+\frac{\gamma N}{4}n(n-1)\expval{\hat J_x^{n-2}}\nonumber\\
&+\frac{\gamma N}2\sum_{p=0\;\text{even}}^{n-4} \binom{n}{p} \expval{\hat J_x^p}\nonumber\\
&-\gamma\sum_{p=0\;\text{even}}^{n-2} \binom{n}{p-1} \expval{\hat J_x^p},\label{full_evolution}
\end{align}
and similarly for $n$ odd we recover Eq.~(\ref{full_evolution}) but with the sums restricted to odd values of $p$. Of the four terms present on the RHS of Eq.~(\ref{full_evolution}), the first two are equivalent to those found through the Fokker-Planck equation (see App.~\ref{det_fp} for details), while the last two terms cause deviations from the Fokker-Planck solution. Applying the usual time-independent Holstein-Primakoff approximation to Eq.~(\ref{full_evolution}) we find for $n$ even that
\begin{align}
    \frac{\mathrm d}{\mathrm dt}\expval{\hat X^n}=&-n\gamma\expval{\hat X^n}+\frac{\gamma}{2}n(n-1)\expval{\hat X^{n-2}}\nonumber\\
    &+\frac{\gamma}{2}\sum_{p=0\;\text{even}}^{n-4}\frac{1}{2^{p/2}N^{(n-p)/2-1}}\binom{n}{p}\expval{\hat X^p}\nonumber\\
    &-\gamma\sum_{p=0\;\text{even}}^{n-2}\frac{1}{2^{p/2}N^{(n-p)/2}}\binom{n}{p-1}\expval{\hat X^p}.\label{full_X_evol}
\end{align}
The last two terms in Eq.~(\ref{full_X_evol}) are always on the order of at most $N^{-1}$. Therefore for sufficiently large $N$ these last two terms are negligible compared to the first two terms, leaving us 
\begin{equation}
    \frac{\mathrm d}{\mathrm dt}\expval{\hat X^n}\approx -n\gamma\expval{\hat X^n}+\frac{\gamma}{2}n(n-1)\expval{\hat X^{n-2}}.\label{fp_X_evol}
\end{equation}
This evolution is identical to the evolution of the moments as governed by the Fokker-Planck equation in Eq.~(\ref{fp}).

One may obtain the time evolution of $\hat J_y^n$ through a similar derivation to the one presented above in Eqs.~(\ref{eq:simplifiedAME}-\ref{fp_X_evol}). While we have here omitted the derivation of the equation of motion for $\hat J_y^n$ due to its similarity with the above derivation for $\hat J_x^n$, it may be obtained as a special case of the derivation presented in Apps.~\ref{generalization} and \ref{evol_jy}, with the choice of optical pumping as local decoherence. The evolution reads
\begin{equation}
    \frac{\mathrm{d}}{\mathrm{d}t}\expval{\hat J_y^n}\approx-n\gamma\expval{\hat J_y^n}+\frac{\gamma N}{4}n(n-1)\expval{\hat J_y^{n-2}}.
\end{equation}
Similarly it follows from Apps.~\ref{generalization} and \ref{evol_jy} that the evolution of any moment of an arbitrary linear combination of $\hat J_x$ and $\hat J_y$ -- and thus a linear combination of the quadratures $\widetilde X$ and $\widetilde P$ -- is identical to Eq.~(\ref{fp_X_evol}). Furthermore, since the inverse Radon transformation uniquely defines a joint probability distribution by its marginals \cite{Radon1996}, we see that Eq.~(\ref{fp}) must be the unique evolution of the Wigner function for an arbitrary initial state in the Fokker-Planck picture.

\begin{figure}
\label{fourth_mom}
\includegraphics[width=1\linewidth]{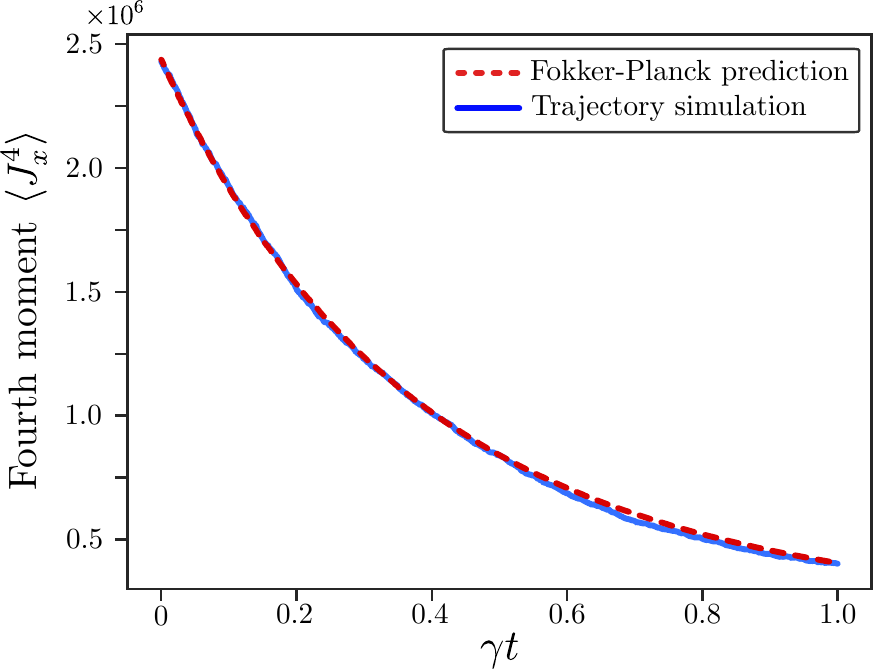}
\caption{Time evolution of the fourth moment, $\langle \hat J_x^4 \rangle$,
for an initial second-excited Dicke state, $\ket{J=N/2, M=N/2-2}$ evolving under local decoherence. The results are shown for $N=1000$ spins. The red curve is the analytic prediction obtained as the solution to Eq.~(\ref{fp_X_evol}) for $n=4$, derived using the Fokker-Planck picture. The blue curve is the average over 4000 quantum trajectories simulated using the method of Ref.~\cite{zhang}. We observe excellent agreement between the two methods for all times considered here, including for long times.}
\label{fig:fourth_mom}
\end{figure}

To investigate numerically the timescales of our analytical results -- that the evolution of moments obeys a Fokker-Planck equation for any initial state -- we compare the prediction in Eq.~(\ref{fp_X_evol}) to simulations conducted using the method of quantum trajectories described in Ref.~\cite{zhang} for a system of $N=1000$ spins. The initial state is chosen to be the second excited Dicke state $\ket{\Psi_i} = \ket{N/2,N/2-2}$. Figure~\ref{fig:fourth_mom} shows a comparison of our analytical results with the results obtained by averaging over 4000 quantum trajectories. The analytic prediction made using Eq.~(\ref{fp_X_evol}) matches that of the simulation very well for all times displayed, surprisingly even for long times~$\gamma t\sim 1$ where the Holstein-Primakoff approximation is no longer valid. We note that the validity of Eq.~(\ref{fp_X_evol}) relies only on the assumption of large $N$, and not the Holstein-Primakoff approximation. This explains why we observe good agreement between the simulation and analytic prediction beyond the Holstein-Primakoff regime.

\subsection{General local decoherence}\label{subsec:genlocaldecoh}
Until now we considered the local decoherence to be optical pumping, described via the local spin-flip raising and lowering operators $\hat\sigma_+^{(i)}$ and $\hat\sigma_-^{(i)}$. We now extend our results to single-channel general local decoherence described by the local operator
\begin{equation}
\hat\ell^{(i)}=\ell_I\mathds1^{(i)}+\ell_+\hat\sigma_+^{(i)}+\ell_-\hat\sigma_-^{(i)}+\ell_z\hat\sigma_z^{(i)}, \label{eq:arbitraryjumpoperator}
\end{equation}
where the coefficients $\ell_q$ are complex numbers and $\{\mathds1^{(i)},\hat\sigma_-^{(i)},\hat\sigma_+^{(i)},\hat\sigma_z^{(i)}\}$ are local operators on the $i^{th}$ site. Generalization to multi-channel local decoherence then follows as a trivial extension. In the following we present the key results, while detailed calculations will be omitted here and instead can be found in App.~\ref{generalization}.

Analogous to Eq.~(\ref{full_evolution}), the evolution of $\expv{\hat J_x^n}$ under an arbitrary local decoherence operator on the form Eq.~(\ref{eq:arbitraryjumpoperator}) is
\begin{align}
\frac{\mathrm d}{\mathrm dt}\expv{\hat J_x^n}\approx&-\frac12n\omega \expv{\hat J_x^n}+\frac N8n(n-1)\omega \expv{\hat J_x^{n-2}}\nonumber\\
&-nN\Re(\ell_z^*(\ell_+-\ell_-))\expv{\hat J_x^{n-1}}\nonumber\\
&+|\phi|\frac n2\expv{\hat J_x^{n-1}}\expval{\hat R_x(\alpha)\hat J_z\hat R_x^\dagger(\alpha)},\label{second_appx}
\end{align}
where $ \omega =4\left(|\ell_z|^2+|\ell_+-\ell_-|^2\right)$, $\hat R_x(\alpha)=\exp[-i\alpha \hat J_x]$, and
\begin{align}
\phi =|\phi|e^{i\alpha}=&\, \ell_z^* (2  \ell_I + \ell_- + \ell_+ )+ \ell_I^* (-2  \ell_z + \ell_- - \ell_+ ) \nonumber\\
&+ \ell_-^* ( \ell_z + \ell_I + \ell_+ )+ \ell_+ ^* ( \ell_z - \ell_I - \ell_- ).
\end{align}
In Eq.~(\ref{second_appx}) we have made the approximation $N \gg n$, which is reasonable for the systems of interest. The final term in Eq.~(\ref{second_appx}) can be solved exactly, as derived in App.~\ref{generalization} and \ref{evol_jy}. Inserting this solution to the final term, Eq.~(\ref{second_appx}) takes the form of a 1D Fokker-Planck equation (see App.~\ref{det_fp}). We therefore see that regardless of the form of local decoherence, the evolution of the moments $\langle \hat J_x^n \rangle$ will always be Fokker-Planck in the limit of large $N$. The drift and diffusion coefficients of this Fokker-Planck equation are 
\begin{align}
    \mu(J_x,t)=&-\frac{\omega}{2}J_x-N\Re(\ell_z^*(\ell_+-\ell_-))\nonumber\\
    &+\frac{|\phi|}{2}\expval{\hat R_x(\alpha)\hat J_z\hat R_x^\dagger (\alpha)}(t),\\
    D(J_x,t)=&\frac{\omega N}{8},
\end{align}
respectively. This follows by noting that the moments of a 1D Fokker-Planck equation evolve according to
\begin{align}
    \frac{\mathrm d}{\mathrm dt}\langle\hat J_x^n\rangle=&n\int\mathrm{d}J_x\, J_x^{n-1}\mu(J_x,t)p_x(J_x,t)\nonumber\\
    &+n(n-1)\int\mathrm{d}J_x\, J_x^{n-2}D(J_x,t)p_x(J_x,t),\label{eq:Jx_fokker_planck}
\end{align}
where $p_x(J_x,t)$ is the marginal distribution
\begin{align}
    p_x(J_x,t)=\Tr[\rho(t)\ketbra{J,m_x=J_x}],\label{eq:Jx_marginal}
\end{align}
and $\ket{J,m_x=J_x}$ is an eigenvector of $\hat J_x$. This shows that the moments of $\hat \rho(t)$ in Eq.~(\ref{eq:Jx_marginal}) evolves according to a Fokker-Planck equation (we derive the form of Eq.~(\ref{eq:Jx_fokker_planck}) in App.~\ref{det_fp}).

\section{Applications to quantum Fisher information}
\label{sec:qfi}
An important application of the preparation of nonclassical states of atomic ensembles is quantum metrology~\cite{Pezze2018}. The quantum advantage one can achieve, however, will be limited by decoherence~\cite{DD}. The quantum Fisher information (QFI) is one such measure of metrological usefulness, and in this section we present a method of approximating the QFI using the thermalizing picture described previously to show how local decoherence affects the system. Our method may in particular be useful when one knows the optimal measurement (POVM) that achieves the Cramer-Rao bound~\cite{Braunstein1994}, and one can then optimize classical Fisher information (CFI) on the bosonic mode associated with that measurement.

We consider the estimation of a parameter $\theta$. The QFI~\cite{QFI2009} can be written as
\begin{align}\label{main_qfi}
    \mathcal F_Q[\hat\rho _0; \hat A]&=2\sum_{\lambda,\lambda'}\frac{(\lambda-\lambda')^2}{\lambda+\lambda'}|\mel{\lambda}{\hat A}{\lambda'}|^2,
\end{align}
where $\hat A$ is the generator of some family of states
\begin{equation}
    \hat\rho_\theta=e^{-i\hat A\theta} \hat\rho_0 e^{i \hat A\theta},
\end{equation}
and $\ket{\lambda}$ is an eigenvector of $\hat\rho _0$ with corresponding eigenvalue $\lambda$. As discussed in Sec.~\ref{sec:optical_pumping} we can treat the collective states as outer products of pure states, hence in the following we drop the double-bar notation for simplicity. Using the Holstein-Primakoff approximation we write that
\begin{align}\label{operator_relationX}
\hat J_x\ket{J,M}&=\sqrt{\langle \hat J\rangle}\hat X\ket{n},\\
\label{operator_relationP}\hat J_y\ket{J,M}&=\sqrt{\langle \hat J\rangle}\hat P\ket{n},
\end{align}
where $\vert J,M \rangle = \sqrt{1/d_N^J}\sum_\alpha \vert J,M,\alpha\rangle$. To make use of Eq.~(\ref{main_qfi}), we need to rewrite it using the relations Eqs.~(\ref{operator_relationX}-\ref{operator_relationP}). To this end we consider the reduced density matrix $\hat\rho_J=\Tr_{\backslash J}(\hat\rho)$, where all irreps of $\hat \rho$ except those with total angular momentum $J$ have been traced out. The eigenvectors and eigenvalues of $\hat\rho_J$ will be denoted as $\ket{\lambda_J}$ and $\lambda_J$ respectively. We assume that the initial state has no coherences between different irreps, which implies that $\hat\rho$ will be block diagonal in the irreps at all times. The eigenvectors $\ket{\lambda}$ are thus the direct extension of the irrep eigenvectors $\ket{\lambda_J}$ for all $J$, with eigenvalues $\{\lambda\} = \{\lambda_J\}$. Inserting this into Eq.~(\ref{main_qfi}) we find that 
\begin{equation}
    \mathcal F_Q[\hat \rho;\hat A]=2\sum_J\sum_{\lambda_J,\lambda_J'}\frac{(\lambda_J-\lambda_J')^2}{\lambda_J+\lambda_J'}|\mel{\lambda_J}{\hat A}{\lambda_J'}|.\label{eq:qfi_over_J}
\end{equation}
In order to relate Eq.~(\ref{eq:qfi_over_J}) to the density operator in the thermalizing bosonic mode, we approximate each irrep as having collective state structure proportional to the Fock state structure of the thermalizing bosonic mode, with the proportionality constant given by the probability to be in that particular irrep, $p_J$. We make this approximation by first letting $\lambda_J=p_J\lambda_{\text{HP}}$, where $\lambda_{\text{HP}}$ are the eigenvalues of the density operator in the thermalizing picture, and further assuming that $\ket{\lambda_J}=\ket{\lambda_\text{HP}}$. Applying these approximations to Eq.~(\ref{eq:qfi_over_J}) and transforming the operator $\hat A$ using Eqs.~(\ref{operator_relationX}-\ref{operator_relationP}) we obtain
\begin{align}
    \mathcal F_Q[\hat\rho;\hat A]=&2\sum_J\sum_{\lambda_{\text{HP}},\lambda_\text{HP}'}p_J\frac{(\lambda_\text{HP}-\lambda_\text{HP}')^2}{\lambda_\text{HP}+\lambda_\text{HP}'}\nonumber\\
    &\;\;\;\;\;\;\;\;\times|\mel{\lambda_\text{HP}}{\langle \hat J\rangle\hat A_{\text{HP}}}{\lambda_\text{HP}'}|\\
    =&2\langle\hat J\rangle\sum_{\lambda_{\text{HP}},\lambda_\text{HP}'}\frac{(\lambda_\text{HP}-\lambda_\text{HP}')^2}{\lambda_\text{HP}+\lambda_\text{HP}'}\nonumber\\
    &\;\;\;\;\;\;\;\;\times |\mel{\lambda_\text{HP}}{\hat A_{\text{HP}}}{\lambda_\text{HP}'}|\\
    =&\frac{N}{2}e^{-2\gamma t}\mathcal F_Q[\hat\rho_{\text{HP}},\hat A_{\text{HP}}].
\end{align}
Equivalently, for a general operator $\hat A^{(k)}$ of order $k$ in $\hat J_x$ and $\hat J_y$ we have
\begin{align}
    \mathcal F_Q[\hat\rho _\text{spin};\hat A_\text{spin}^{(k)}]\approx\left(\frac N2\right)^{k}e^{-2k\gamma t}\mathcal F_Q[\hat\rho _\text{HP};\hat A_\text{HP}^{(k)}],\label{qfi_appx}
\end{align}
where $\hat A_{\text{HP}}^{(k)}$ is equal to $\hat A^{(k)}$ with $\hat J_x\to\hat X$ and $\hat J_y\to\hat P$. The density operator $\hat\rho_{\text{HP}}$ is equivalent to the Wigner function in the thermalizing picture, and thus Eq.~(\ref{qfi_appx}) provides a mapping between the QFI of the spin system and the QFI of the thermalizing bosonic mode.

\begin{figure}
\includegraphics[width=1\linewidth]{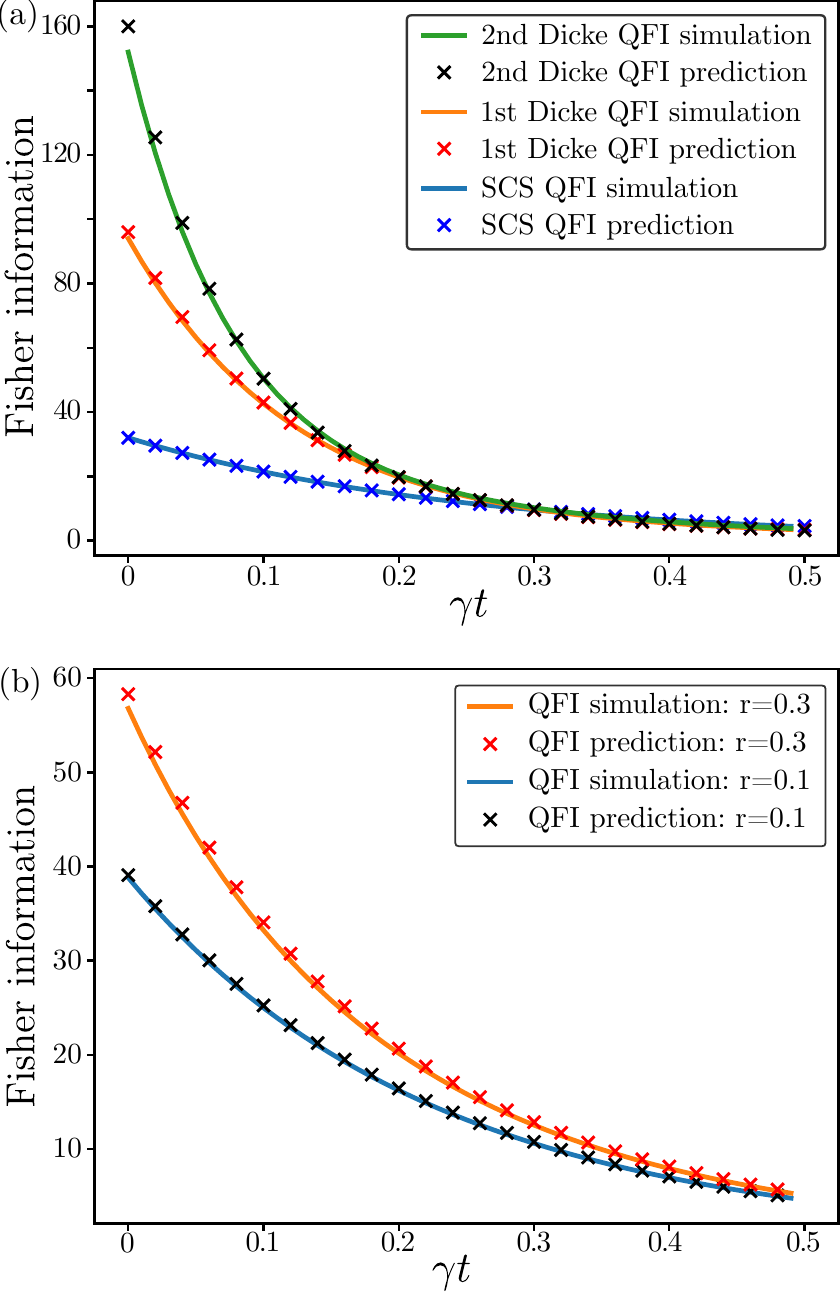}
\caption{Quantum Fisher Information (QFI) for different initial states with $N=32$ spins, evolving under local decoherence. The QFI corresponds estimating a rotation about the $\hat J_x$ axis, plotted a function of the initial state has evolved under optical pumping, $\gamma t$. Solid lines denote the QFI obtained by direct simulations of the Lindblad master equation, while crosses are given by the QFI obtained using Eq.~(\ref{qfi_appx}). (a) QFI for three initial Dicke states: the spin coherent state (SCS) $\ketbra{0}$, the first Dicke state $\ketbra{1}{1}$, and the second Dicke state $\ketbra{2}$. (b) QFI for initial spin squeezed states characterized by the squeezing parameter $r$. For both initial Dicke states as well as initial spin squeezed states we observe great agreement between the simulated QFI and the QFI obtained using our formalism [Eq.~(\ref{qfi_appx})]. The observed deviations at early times are a result of finite size effects, and they disappear for larger $N$.}
\label{fig:qfi_dicke}
\end{figure}

  To demonstrate the utility of Eq.~(\ref{qfi_appx}) we first consider the case where $\hat A = \hat J_x$, corresponding to a rotation about the $\hat J_x$-axis (and equivalently for the bosonic mode a translation in the $\hat P$ direction). For this observable Eq.~(\ref{qfi_appx}) reveals that the QFI of the spin system is proportional to the QFI of the thermalizing Wigner function with proportionality constant $(N/2) \exp(-2 \gamma t)$. When considering the initial state to be a single low-excitation Dicke state, one finds the QFI to be well captured by Eq.~(\ref{qfi_appx}). This is illustrated in Fig.~\ref{fig:qfi_dicke}(a), where the exact QFI found by simulating the spin system is compared to the approximation Eq.~(\ref{qfi_appx}) obtained using the thermalizing picture of the bosonic mode. Strong agreement is observed between the the simulated and predicted QFI, with slight deviations at short times due to finite size effects that disappear for larger system sizes $N$.

Figure~\ref{fig:qfi_dicke}(b) further demonstrates the utility of Eq.~(\ref{qfi_appx}) for the case of initial spin squeezed states via a two-axis twisting unitary \cite{Kitagawa1993}
\begin{equation}
    \hat U_{\text{TAT}}=\exp[-\frac{\chi}{2}\left(\hat J_+^2-\hat J_-^2\right)]
\end{equation}
which, unlike the single Dicke states considered previously, have finite coherences in the collective state basis that decay over time. For different values of the squeezing parameter $r=\chi N$ we again see excellent agreement between the simulated and predicted QFIs. Together, these two examples demonstrate the utility of Eq.~(\ref{qfi_appx}) when calculating the QFI of spin states using the thermalizing picture. This approximation is particularly useful when the bosonic space is more convenient to work with than the collective state space, or when the optimal POVM in the thermalizing picture is known.

\section{Conclusion}\label{sec:conclusion}
In this article we have used the formalism of collective states to show that the evolution of many-body spin systems undergoing local decoherence can be represented as a bosonic mode evolving according to either a diffusion equation or the Fokker-Planck equation. We have established the connection between local decoherence and Fokker-Planck evolution by showing that, in the limit of large system sizes $N$, the moments of the $\hat J_x$ and $\hat J_y$ distributions evolve identically to those of $\hat X$ and $\hat P$ under the Fokker-Planck equation. Using this formalism one can obtain all moments of products of $\hat J_x$ and $\hat J_y$, but not necessarily the microstates of the system.

We have shown that there are two useful ``pictures'' in which to represent local decoherence using a bosonic mode. The thermalizing picture is represented by a diffusion equation, while the Fokker-Planck picture is represented by a Fokker-Planck equation. Each picture places some time dependence into the operators, and each picture has its own unique advantages in calculating physical quantities of interest.

We saw in Sec.~\ref{sec:qfi} that the thermalizing picture is particularly useful when deriving analytic approximations to certain quantities of interest, such as the quantum Fisher information or marginals of the Wigner function. This arises from the fact that the thermalizing picture is obtained through the Holstein-Primakoff approximation where angular momentum states $\ket{J,M}$ are mapped to bosonic Fock states $\ket{n=J-M}$. As discussed in Sec.~\ref{sec:qfi}, this mapping approximately preserves the Hilbert space structure of the state in the limit of large $N$. Finally, the diffusion equation is often easier to solve analytically than the Fokker-Planck equation due to the absence of a drift term in the former.

On the other hand, the marginal distributions of the Wigner function in the thermalizing picture are not proportional to the probability distributions for $\hat J_x$ and $\hat J_y$. To obtain this proportionality one needs to switch to the Fokker-Planck picture, which adds more time dependence into the operators in exchange for the correct marginal distributions. In this picture the Lindblad master equation is identical to that of a damped harmonic oscillator, and under this master equation all initial states evolve towards the vacuum, reducing the size of the Hilbert space one must retain. This is in stark contrast to the thermalizing picture, which spreads amplitude over the Fock space as time increases. As a result one may wish to use the Fokker-Planck picture when simulating the state in Hilbert space, as the dimension of the space one must retain to obtain an accurate approximation shrinks as optical pumping takes effect.

Careful consideration should be made in deciding how long to describe system behavior when the evolution of the system is governed by collective dynamics in addition to local decoherence. We found that states evolving under both collective and local decoherence require larger system sizes $N\sim100$ to converge to predicted values on time scales $\gamma t\sim1$. This is in contrast to systems where only local decoherence is present, which appear to converge to predicted values for moderate system sizes $N\sim16$, even for long times $\gamma t\sim1$.

The formalism presented in this work assumed uniform coupling of the atoms to the probe mode of the light. This assumption will only be approximately true in certain situations, and thus a generalization to non-uniform coupling is necessary to extend our present work to the dynamics of more general systems. Ideally, such extensions could be done perturbatively through adjustments to the equation of motion of the Wigner function.

We wish to highlight that the Fokker-Planck equation emerged from an Ornstein–Uhlenbeck process. There is one-to-one correspondence between a Fokker-Planck equation and an ensemble of a stochastic differential equation including the drift term and random walks (or diffusion). In principle our result implies that local operators can be treated as stochastic ``kicks'' to the system, encapsulated in a stochastic differential equation.

Our extension of the Chase and Geremia formalism~\cite{chase} to a bosonic mode description not only provides physical intuition about the system dynamics, but also provides practical utility in simulating nonclassical dynamics of large spin ensembles.  This is particularly difficult for nonGaussian states described by more than a few moments of the distribution. The mapping to a bosonic mode allows us to make use of the full tool box of continuous-variable quantum mechanics, including visualization in phase space.

\section{Acknowledgments}
The authors thank F. Elohim Becerra and Jacob S. Nelson for helpful discussions, and Marco A. Rodr\'{i}guez-Garc\'{i}a for insights into the quantum Fisher information of bosonic systems. This work is supported by funding from the National Science Foundation, Grants PHY-2011582, PHY-2116246, and the NSF Quantum Leap Challenge Institutes program, Award No. 2016244.

\appendix

\section{Collective states can be treated as an outer products of pure states}
\label{sec:outer_product_proof}
While the collective states $\col{J,M}{J,M'}$ were initially defined as statistical mixtures of pure states in Sec.~\ref{sec:optical_pumping}, in this Appendix we will show that in certain instances it is valid to treat $\col{J,M}{J,M'}$ as an outer product $\ketbra{J,M}{J,M'}$ of two pure states $\ket{J,M}$ and $\ket{J,M'}$. Specifically, this treatment is valid when one wishes to describe expectation values of collective operators.

Recall that the collective states are defined as
\begin{align}
    \col{J,M,N}{J,M',N}=\frac{1}{d^J_N} \sum_\alpha\ketbra{J,M,\alpha}{J,M',\alpha},
\end{align}
and that the relationship between the collective state $\col{J,M,N}{J,M^\prime,N}$ and the corresponding outer product of pure states reads
\begin{align}\label{app:eq:coll_states}
    &\frac1{d_N^J}\sum_{\alpha,\alpha'}\ketbra{J,M,\alpha}{J,M',\alpha'} \nonumber\\
    &=\col{J,M,N}{J,M',N}+\frac1{d_N^J}\sum_{\alpha\neq\alpha'}\ketbra{J,M,\alpha}{J,M',\alpha'}.
\end{align}
To show that $\col{J,M,N}{J,M',N}$ may be treated as the outer product of pure states, we need to show that the second term on the RHS of Eq.~(\ref{app:eq:coll_states}) does not contribute to expectation values of any observable quantity.

To this end let us first consider the action of a local operator $\hat\ell = \sum_\lambda\hat\ell^{(\lambda)}$ on a Dicke state:
\begin{align}
    \sum_{\lambda=1}^N\hat\ell^{(\lambda)}\ket{J',M',\alpha'}=\sum_{J,M,\alpha}c_{J,M,\alpha}\ket{J,M,\alpha},
\end{align}
where $\hat\ell^{(\lambda)}$ is some arbitrary operator on the $i$th site. In general the coefficient $c_{J,M,\alpha}$ will depend on the degeneracy quantum number $\alpha$ in addition to the usual quantum numbers $J$ and $M$. However, the question is whether these degeneracies contribute to any observable quantity. The degeneracies will not contribute to any observable quantity if for $\alpha \neq \alpha'$ we have the following condition,
\begin{equation}\label{hypoth}
    \Tr(\hat C^{(k)}\sum_{\lambda=1}^N \hat\ell^{(\lambda)}\ketbra{J,M,\alpha}{J',M',\alpha'}\hat\ell^{(\lambda) \dagger})=0,
\end{equation}
for any operator $\hat C^{(k)}$ defined by
\begin{equation}
    \hat C^{(k)}=\hat C_1 \hat C_2\dots \hat C_k=\sum_{i_1,i_2,\dots,i_k}^N \hat c_1^{(i_1)} \hat c_2^{(i_2)}\dots \hat c_k^{(i_k)}. \label{app:eq:collectiveoperator}
\end{equation}
and any local operator $\hat\ell = \sum_\lambda \ell^{(\lambda)}$. We note that the operator $\hat C^{(k)}$ is a product of $k$ collective operators, where each collective operator is of the form
\begin{align}
    \hat C_{j}=\sum_{i_{j}=1}^N \hat c_{j}^{(i_{j})}. \label{app:eq:Cexpansion}
\end{align}
Intuitively, $\hat C^{(k)}$ may be any product of collective spin operators and identities, with rank $k$. All operators of interest for our studies in the main text are on the form of Eq.~(\ref{app:eq:collectiveoperator}). Therefore it is sufficient to show that the LHS of Eq.~(\ref{hypoth}) is zero for such operators when $\alpha\neq\alpha'$ to establish the desired result. 

To prove Eq.~(\ref{hypoth}) we first need to establish the following result. Consider a product of $k$ collective operators $\hat C^{(k)}$, and let $\hat\ell=\sum_\lambda \hat\ell^{(\lambda)}$ and $\hat\mu = \sum_\lambda \hat\mu^{(\lambda)}$ be local operators. We then have
\begin{equation}
    \sum_{\lambda} \hat\ell^{(\lambda) \dagger} \hat C^{(k)} \hat\mu^{(\lambda)} = \sum_m \alpha_m \, \hat D_m^{(k_m)}, \label{app:eq:inductionassumption}
\end{equation}
where $\hat D_m^{(k_m)}$ is a product of $k_m \leq k+1$ collective operators and $\alpha_m$ is a complex number. The result states that applying the two local operators $\hat\ell^{(\lambda)\dagger}$ and $\hat\mu^{(\lambda)}$ to the product of $k$ collective operators $\hat C^{(k)}$ and then taking the sum over $\lambda$ yields a sum over products of collective operators $\hat C_m^{(k_m)}$ whose ranks are at most $k_m \leq k+1$.

We will show Eq.~(\ref{app:eq:inductionassumption}) using an induction proof. First we show the case $k=1$, which reads
\begin{align}
    \sum_\lambda \hat\ell^{(\lambda)\dagger} \hat C^{(1)} \hat\mu^{(\lambda)}=&\sum_{\lambda,i} \hat\ell^{(\lambda) \dagger}\,\hat c^{(i)}\hat\mu^{(\lambda)} \nonumber\\
    =&\sum_{\lambda,i} \hat\ell^{(\lambda) \dagger} \hat\mu^{(\lambda)} \hat c^{(i)} +\sum_\lambda \hat\ell^{(\lambda) \dagger}\,[\hat c^{(\lambda)},\hat\ell^{(\lambda)}]\nonumber\\
    =&\hat A \hat C + \hat B. \label{app:eq:k1result}
\end{align}
Here we have introduced the new operators
\begin{align}
    \hat A& = \sum_{\lambda} \hat\ell^{(\lambda) \dagger} \hat\mu^{(\lambda)}, \label{app:eq:defAk1}\\
    \hat B& = \sum_{\lambda} \hat\ell^{(\lambda) \dagger} [\hat c^{(\lambda)}, \hat\mu^{(\lambda)}] \equiv \sum_\lambda \hat\ell^{(\lambda) \dagger}\; \tildehat{\mu}^{(\lambda)}, \label{app:eq:defBk1}
\end{align}
where we in the last equality have defined $\tildehat{\mu}^{(\lambda)}=[\hat c^{(\lambda)},\hat\mu^{(\lambda)}]$, which also is a local operator. It follows directly from Eqs.~(\ref{app:eq:defAk1}-\ref{app:eq:defBk1}) that $\hat A$ and $\hat B$ are collective operators, and thus Eq.~(\ref{app:eq:k1result}) is the sum of products of collective operators of at most rank 2, which we can rewrite as
\begin{equation}
    \sum_\lambda \hat\ell^{(\lambda)\dagger} \hat C^{(1)} \hat\mu^{(\lambda)} = \sum_m \alpha_m \hat D_m^{(k_m)},
\end{equation}
where $k_m \leq 2$. This shows Eq.~(\ref{app:eq:inductionassumption}) for the case $k=1$.

We now make the induction step by assuming that Eq.~(\ref{app:eq:inductionassumption}) holds for products of $k$ collective operators $\hat C^{(k)}$ and showing that it then also holds for products of $k+1$ collective operators $\hat C^{(k+1)}$. We write
\begin{align}
    \sum_\lambda \hat\ell^{(\lambda)\dagger} \hat C^{(k+1)} \hat\mu^{(\lambda)} =& \sum_\lambda \hat\ell^{(\lambda)\dagger} \hat C_1 \ldots \hat C_k \hat C_{k+1} \hat\mu^{(\lambda)} \nonumber\\
    =& \sum_\lambda \hat\ell^{(\lambda)\dagger} \hat C_1\ldots \hat C_k \, \hat\mu^{(\lambda)} \hat C_{k+1} \nonumber\\
    &+ \sum_\lambda \hat\ell^{(\lambda)\dagger} \hat C_1\ldots \hat C_k \,\tildehat{\mu}^{(\lambda)}, \label{app:eq:inductionstep}
\end{align}
where the local operator $\tildehat{\mu}^{(\lambda)} = [\hat C_{k+1},\hat\mu^{(\lambda)}]$ is the commutator between $\hat C_{k+1}$ and $\hat\mu^{(\lambda)}$. We can now apply the induction assumption Eq.~(\ref{app:eq:inductionassumption}) to both terms in Eq.~(\ref{app:eq:inductionstep}) and find that each term yields a sum over products of collective operators of rank at most $k+1$. Hence we can write
\begin{equation}
    \sum_\lambda \hat\ell^{(\lambda)\dagger} \hat C^{(k+1)} \hat\mu^{(\lambda)} = \sum_n \beta_n \hat D_n^{(k_n)},
\end{equation}
with $\hat D_n^{(k_n)}$ products of collective operators of rank at most $k_n \leq k+1$ and $\beta_n$ complex coefficients. This form is exactly that of Eq.~(\ref{app:eq:inductionassumption}) for $k+1$, hence completing our induction proof of Eq.~(\ref{app:eq:inductionassumption}).

We are now ready to prove the result in Eq.~(\ref{hypoth}), namely that the trace in Eq.~(\ref{hypoth}) is zero for $\alpha \neq \alpha^\prime$ for any operator $\hat C^{(k)}$ defined as in Eq.~(\ref{app:eq:collectiveoperator}). To prove this result we first note that we can use the cyclic property to write $\Tr[(\sum_{\lambda=1}^N \hat\ell^{(\lambda) \dagger} \hat C^{(k)} \hat \ell^{(\lambda)}) \ketbra{J,M,\alpha}{J',M',\alpha'}]$, and then use Eq.~(\ref{app:eq:inductionassumption}) to express the parenthesis $(\sum_{\lambda=1}^N \hat\ell^{(\lambda) \dagger} \hat C^{(k)} \hat\ell^{(\lambda)})$ in terms of sums over products of collective operators. All collective operators as defined by Eq.~(\ref{app:eq:Cexpansion}) can be written as
\begin{align}
    \sum_i \hat c^{(i)}=c_I\hat{\mathds1}+c_+\hat J_++c_-\hat J_-+c_z\hat J_z,\label{eq:collective_operator_expanded}
\end{align}
for complex numbers $c_I$, $c_+$, $c_-$, and $c_z$. We note that the operators on the RHS of Eq.~(\ref{eq:collective_operator_expanded}) do not couple subspaces of different degenerate quantum number $\alpha$, and hence collective operators do not couple degenerate subspaces. Therefore the terms $(\sum_{\lambda=1}^N \hat\ell^{(\lambda) \dagger} \hat C^{(k)} \hat\ell^{(\lambda)})$ will be diagonal in $\alpha$, with the implication that Eq.~(\ref{hypoth}) must be zero for $\alpha\neq \alpha'$. We thus find that the off-diagonal (in $\alpha$) terms in Eq.~(\ref{app:eq:coll_states}) do not contribute to any expectation value for collective operators (or products of these), hence we may treat $\col{J,M,N}{J,M',N}$ as the outer product of pure states.

\section{Derivation of Fock state populations}
\label{sec:p_n}
In Sec.~\ref{sec:optical_pumping} we derived a solution to the collective state populations $p_{J,M}(t)$ for an initial spin coherent state, and then wrote an approximation to this state in a bosonic mode using the Holstein-Primakoff approximation. The Fock state populations $p_n(t)$ in Eq.~(\ref{eq:NlimitHP}) were found by calculating 
\begin{equation}
    p_n(t)=\lim_{N\to\infty}\sum_J p_{J,M=J-n}(t), \label{app:eq:fockstatepop}.
\end{equation}
In this Appendix we derive the result in detail and explicitly state the utilized approximations.

We begin by noting that from Eq.~(\ref{pjms}) it follows that
\begin{equation}
    p_{J,M}(t)=p_{J,J}(t)\tanh^{J-M}(\gamma t), \label{app:eq:JminusMForm}
\end{equation}
and also that
\begin{align}
    p_{J,J}(t)=&d^N_J\left(\frac{1+e^{-2\gamma t}}{2}\right)^{\frac N2+J}\left(\frac{1-e^{-2\gamma t}}{2}\right)^{\frac N2-J} \nonumber\\
    =&\frac{4J+2}{N+2J+2}\binom{N}{\frac N2+J} \nonumber\\
    &\hspace{1cm}\left(\frac{1+e^{-2\gamma t}}{2}\right)^{\frac N2+J}\left(\frac{1-e^{-2\gamma t}}{2}\right)^{\frac N2-J}. \label{p_jj}
\end{align}
Equation~(\ref{p_jj}) can be further approximated using the normal approximation, i.e.,
\begin{equation}
    \binom{n}{x}p^xq^{n-x}\approx\frac1{\sqrt{2\pi npq}}e^{-(x-np)^2/2npq}
\end{equation}
for large $n$, $np$, and $nq$. This approximation holds when $N$ is large and enough time has passed such that
\begin{equation} \label{app:eq:timeconstraint}
    N\left(\frac{1-e^{-2\gamma t}}{2}\right)\gg1.
\end{equation}
The larger $N$ is, the shorter the time $\gamma t$ needs to be. Assuming that $\gamma t$ satisfies the condition Eq.~(\ref{app:eq:timeconstraint}), we apply the normal approximation to find that
\begin{align}
    \nonumber p_{J,J}(t)\approx&\frac{4J+2}{2J+2+N}\sqrt{\frac{2}{\pi N(1-e^{-4t})}}\\
    &\times\exp[-\frac{(J-\frac N2e^{-2t})^2}{\frac N2(1-e^{-4t})}]. \label{app:eq:pJJapprox}
\end{align}

We are now in a position to calculate $p_n(t)$ in Eq.~(\ref{app:eq:fockstatepop}). Let us start by considering $p_0(t) = \lim_{N\rightarrow\infty}\sum_J p_{J,J}(t)$, where $p_{J,J}(t)$ is given in Eq.~(\ref{app:eq:pJJapprox}). We make two observations. First, the approximation obtained for $p_{J,J}$ contains a Gaussian whose standard deviation scales with $\sqrt N$, and hence this distribution is sharply peaked around its mean for large $N$. Second, we note that for large $N$ we can approximate the sum over $J$ in $p_0$ as an integral over $J$. These two observations allow us to rewrite the sum as
\begin{align}
    \nonumber\sum_Jp_{J,J}\approx&\frac{4\bar J+2}{2\bar J+2+N}\sqrt{\frac{2}{\pi N(1-e^{-4\gamma t})}}\\
    &\times\int_{0}^{\infty}\mathrm dJ\;\exp[-\frac{(J-\frac N2e^{-2\gamma t})^2}{\frac N2(1-e^{-4\gamma t})}],
\end{align}
where $\bar J\approx\frac N2e^{-2\gamma t}$ is the mean $J$-value of the Gaussian. Evaluating the integral and taking the limit $N\to\infty$ we find that
\begin{equation}
    p_0(t) = \lim_{N\rightarrow\infty}\sum_J p_{J,J}(t)\approx1-\tanh(\gamma t).
\end{equation}
Similarly, for $n>0$ we calculate $p_n(t)$ using Eq.~(\ref{app:eq:fockstatepop}) and find
\begin{align}
    p_n(t) =& \lim_{N\rightarrow\infty} \sum_J p_{J,J-n}(t) \nonumber\\  
    =& \lim_{N\to\infty} \sum_J p_{J,J}(t) \tanh^n(\gamma t) \nonumber\\
    \approx&\tanh^n(\gamma t)(1-\tanh(\gamma t)),
\end{align}
where in the second equality we have used Eq.~(\ref{app:eq:JminusMForm}) with $n=J-M$, and in the final line have used the form of $p_0(t)$ derived above.

As a final comment we argue that the probability distribution of $J$ is proportional to $p_{J,J}$ when the initial state is polarized along $J_z$ so that the Holstein-Primakoff approximation applies. To show this we note that the probability $P_J$ to be in a subspace labeled by $J$ is
\begin{align}
    \mathrm{P}_J=&p_{J,J}\sum_{M=-J}^J\tanh^{J-M}(\gamma t) \nonumber\\
    =&p_{J,J}\frac{\tanh^{2J+1}(\gamma t)-1}{\tanh(\gamma t)-1}.\label{total_j_pop}
\end{align}
For sufficiently large $J$ the fraction in Eq.~(\ref{total_j_pop}) is approximately constant in $J$, and therefore $\mathrm P_J\propto p_{J,J}$. It thus follows that the mean value of $J$ is
\begin{equation}
    \bar J\approx \frac N2e^{-2\gamma t}.
\end{equation}

\section{Determining when time evolution is governed by the Fokker-Planck equation}
\label{det_fp}
A central goal of this paper is to show whether or not a particular equation of motion can be written as a Fokker-Planck equation. Because our work focuses on deriving equations of motion for the moments $\hat J_x^n$, in this Appendix we will look at the evolution of such moments under a general 1D Fokker-Planck equation.

The general form of a 1D Fokker-Planck equation for probability density function $p(x,t)$ is
\begin{equation}\label{1d_fp}
\party{}{t}p(x,t)=-\party{}{x}\mu(x,t)p(x,t)+\party{^2}{x^2}D(x,t)p(x,t),
\end{equation}
where $\mu(x,t)$ is the ``drift function'' and $D(x,t)$ is the ``diffusion function''. The evolution of the moments of a Fokker-Planck equation can be found by integrating Eq.~(\ref{1d_fp}) with $x^n$
\begin{align}
\party{}{t}\expv{x^n}=&-\int \mathrm{d}x\, x^n \party{}{x}\mu(x,t)p(x,t)\nonumber\\
&+\int\mathrm{d}x\, x^n\party{^2}{x^2}D(x,t)p(x,t) \nonumber\\
=&n\int\mathrm{d}x\, x^{n-1}\mu(x,t)p(x,t)\nonumber\\
&+n(n-1)\int\mathrm{d}x\, x^{n-2}D(x,t)p(x,t). \label{mom_evol2}
\end{align}
Equation~(\ref{mom_evol2}) implies that any moments evolving according to a Fokker-Planck equation will have two terms in their equation of motion, one proportional to $n$ and the other proportional to $n(n-1)$. In the case that $\mu\propto x$ and $D$ is constant, the integrals in Eq.~(\ref{mom_evol2}) reduce to expectation values of $x^{n}$ and $x^{n-1}$, which is the form of Eq.~(\ref{fp_X_evol}).

We will use Eq.~(\ref{mom_evol2}) to identify Fokker-Planck equations when the coefficients $\mu$ and $D$ are easily identifiable. We find that this is possible when deriving the equations of motion for optical pumping, and even for generalized local decoherence.

\section{Generalizing the formalism to other forms of local decoherence}
\label{generalization}
Section~\ref{sec:rig} presented analytical results for local optical pumping and arbitrary initial states, and in Sec.~\ref{subsec:genlocaldecoh} we studied the extension of our analytical results to general local decoherence.  In this Appendix we provide the detailed derivation to show the results of Sec.~\ref{subsec:genlocaldecoh}, namely that the time evolution remains Fokker-Planck under the general form of decoherence specified by the local operator 
\begin{equation}
\hat\ell^{(i)}=\ell_I{\mathds1}^{(i)}+\ell_+\hat\sigma_+^{(i)}+\ell_-\hat\sigma_-^{(i)}+\ell_z\hat\sigma_z^{(i)}. \label{app:eq:generaldecohoper}
\end{equation}
Here $\hat\sigma_+^{(i)}$, $\hat\sigma_-^{(i)}$, and $\hat\sigma_z^{(i)}$ are the raising, lowering, and Pauli $z$ operator on $i$th site, respectively, and the $\ell_i$ coefficients are complex numbers.

To prove this result we consider the jump operators defined in Eq.~(\ref{eq:JumpExpansionGenSpherical}), which we repeat here for convenience as
\begin{equation}\label{spherical_tensor}
\hat L_{j,q}=\sum_J\Lambda_{j,q}(J,N)\;\hat T_{1,q}^{J+j,J}.
\end{equation}
The operators $\hat T_{1,q}^{J+j,J}$ are the generalized spherical tensor operators introduced in Sec.~\ref{sec:rig} and are given by
\begin{equation}
\hat T_{k,q}^{J',J}=\sqrt{\frac{2k+1}{2J'+1}}\sum_{M,M'}C_{k,q,J,M}^{J',M'}\col{J',M'}{J,M},
\end{equation}
where as in the main text $C_{k,q,J,M}^{J',M'}=\braket{J',M'}{k,q,J,M}$ are the Clebsch-Gordan coefficients. The coefficients $\Lambda_{j,q}(J,M)$ are obtained using the formalism in Ref.~\cite{chase} and are given by
\begin{subequations}
\begin{align}
\Lambda_{1,q}(J,M)&=-\widetilde q\sqrt{\frac{(N-2J)(2J+3)}{6}},\\
\Lambda_{0,q}(J,M)&=\widetilde q\sqrt{\frac{(N+2)(2J+1)}{6}},\\
\Lambda_{-1,q}(J,M)&=\widetilde q\sqrt{\frac{(N+2J+2)(2J-1)}{6}},
\end{align}
\end{subequations}
with the parameter
\begin{equation}
\widetilde q=
\begin{cases}
q, & \;\; q=\pm1\\
-\sqrt2, & \;\; q=0
\end{cases}.
\end{equation}
Equation~(\ref{spherical_tensor}) defines the six $\hat L_{j, q}$ operators from Sec.~\ref{sec:rig} (for $q = \pm 1$), plus three additional operators representing local $\hat\sigma_z$ jumps (when $q = 0$).

To prove that local decoherence given by the operator Eq.~(\ref{app:eq:generaldecohoper}) leads to Fokker-Planck evolution, we proceed along similar lines to Sec.~\ref{sec:rig} and consider the evolution of $\hat J_x^n$ under the adjoint master equation. The adjoint master equation for the $n$th moment $\hat{J}_x^n$ reads
\begin{equation}
    \frac{d}{dt} \hat J_x^n=\sum_{j,q} \hat L^\dagger_{j,q}\, \hat J_x^n\, \hat L_{j,q} - \frac{1}{2} \left\{\hat L^\dagger_{j,q}\, \hat L_{j,q},\,\hat J_x^n\right\}, \label{app:eq:AME}
\end{equation}
and inserting the jump operators from Eq.~(\ref{spherical_tensor}) the adjoint master equation takes the form
\begin{widetext}
\begin{align}
\frac{d}{dt} \hat J_x^n =&-\frac{1}{2} \hat A \hat J_x^n-\frac12\hat J_x^n\hat A+|\ell_I|^2N\hat J_x^n+\sum_{r\in\{+,-\}}\left(\ell_r\ell_I^*\hat J_x^n \hat J_r+\ell_I\ell_r^* \hat J_r^\dagger \hat J_x^n\right) \nonumber\\
&+2\ell_z\ell_I^*\hat J_x^n\hat J_z+2\ell_I\ell_z^*\hat J_z \hat J_x^n+\sum_{J,j,q,q'}\ell_{q}^*\ell_{q'}\hat L_{j,q}^\dagger \hat J_x^n\hat L_{j,q'}, \label{jx_evol}
\end{align}
where we have defined the operator
\begin{align}
\hat A=&\left(\frac12|\ell_-|^2+\frac12|\ell_+|^2+|\ell_I|^2+|\ell_z|^2\right)N\mathds1+\left(\ell_-^*\ell_I-\ell_-^*\ell_z+\ell_I^*\ell_++\ell_z^*\ell_+\right)\hat J_+ \nonumber\\
&+\left(\ell_I^*\ell_-+\ell_+^*\ell_I+\ell_+^*\ell_z-\ell_z^*\ell_-\right)\hat J_-+\left(|\ell_-|^2-|\ell_+|^2+2\ell_I^*\ell_z+2\ell_z^*\ell_I\right)\hat J_z \nonumber\\
=&\alpha_IN\mathds1+\alpha_+\hat J_++\alpha_-\hat J_-+\alpha_z\hat J_z. \label{jx_evoleq2}
\end{align}
Here $\hat J_+$ and $\hat J_-$ are the collective spin raising and lowering operators. In Eq.~(\ref{jx_evoleq2}) we have defined the coefficients $\alpha_i$ for notational convenience. In Eq.~(\ref{jx_evol}) the first two terms $-\frac{1}{2} \hat A \hat J_x^n-\frac12\hat J_x^n\hat A$ are due to the anticommutator in the adjoint master equation [Eq.~(\ref{app:eq:AME})]. The third term $|\ell_I|^2N\hat J_x^n$ is the jump-term for the identity part of $\hat L_{j,q}$, while terms four through six $\sum_{r\in\{+,-\}}\left(\ell_r\ell_I^*\hat J_x^n \hat J_r+\ell_I\ell_r^* \hat J_r^\dagger \hat J_x^n\right)+2\ell_z\ell_I^*\hat J_x^n\hat J_z+2\ell_I\ell_z^*\hat J_z \hat J_x^n$ are crossterms between identity and non-identity parts of $\hat L_{j,q}$. Finally, the last term in Eq.~(\ref{jx_evol}) $\sum_{J,j,q,q'}\ell_{q}^*\ell_{q'}\hat L_{j,q}^\dagger \hat J_x^n\hat L_{j,q'}$ is the jump terms involving only non-identity parts of $\hat L_{j,q}$.

Next we apply the rotation $\hat{\mathcal{R}} := \hat R_y(-\pi/2)=\text{exp}(i\pi\hat{J}_y/2)$ to both sides of Eq.~(\ref{jx_evol}) to align $\hat{J}_x^n$ with the $z$-axis. This is done in order to easily represent $\hat{J}_x$ to an arbitrary power, since it is now aligned with the former $\hat{J}_z$ axis for which the collective states form a good basis. Under this rotation the term $\hat A \hat J_x^n$ in Eq.~(\ref{jx_evol}) takes the form
\begin{equation}
\hat{\mathcal{R}}\, \hat A\hat J_x^n\, \hat{\mathcal{R}}^\dagger = \hat R_y\hat A\hat R_y^\dagger \hat J_z^n=\left[\alpha_IN\mathds1+\frac12\left(\alpha_+-\alpha_--\alpha_z\right)\hat J_+ +\frac12\left(\alpha_--\alpha_+-\alpha_z\right)\hat J_-+\left(\alpha_++\alpha_-\right)\hat J_z\right] \hat J_z^n, \label{rotated_A}
\end{equation}
and we find a similar result for the term $\hat J_x^n \hat A$. The jump terms in Eq.~(\ref{jx_evol}) involving crossterms between identity and non-identity jumps becomes
\begin{align}
\hat{\mathcal{R}}\Bigg[\sum_{r\in\{+,-\}}\left(\ell_r\ell_I^*\hat J_x^n \hat J_r+\ell_I\ell_r^* \hat J_r^\dagger \hat J_x^n\right)&+2\ell_z\ell_I^*\hat J_x^n\hat J_z+2\ell_I\ell_z^*\hat J_z \hat J_x^n\Bigg]\hat{\mathcal{R}}^\dagger \nonumber\\
=& \ell_+\ell_I^*\hat J_z^n\left(\frac12\hat J_+-\frac12\hat J_-+\hat J_z\right) +\ell_I\ell_+^*\left(\frac12\hat J_--\frac12\hat J_++\hat J_z\right)\hat J_z^n \nonumber\\
&+\ell_-\ell_I^*\hat J_z^n\left(\frac12\hat J_--\frac12\hat J_++\hat J_z\right) +\ell_I\ell_-^*\left(\frac12\hat J_+-\frac12\hat J_-+\hat J_z\right)\hat J_z^n\nonumber\\
&-2\ell_z\ell_I^*\hat J_z^n\hat J_x-2\ell_I\ell_z^*\hat J_x\hat J_z^n.\label{rotated_I}
\end{align}
Equations \ref{rotated_A} and \ref{rotated_I} provide rotated versions of all terms in Eq.~(\ref{jx_evol}) except for the final term, which we will treat shortly. Before doing so we note that in the $\ket{J,M_z}$ basis the sum of Eqs.~\ref{rotated_A} and \ref{rotated_I} have only diagonal and one-off diagonal elements. The diagonal elements of all rotated terms in Eqs.~(\ref{rotated_A}-\ref{rotated_I}) read
\begin{align}
\bra{J,M}&\hat{\mathcal{R}}\left(-\frac12\hat A\hat J_x^n -\frac12 \hat J_x^n\hat A +\vert \ell_I\vert^2 N \hat{J}_x^n +\sum_{r\in\{+,-\}}\left(\ell_r\ell_I^*\hat J_x^n \hat J_r+\ell_I\ell_r^* \hat J_r^\dagger \hat J_x^n\right)+2\ell_z\ell_I^*\hat J_x^n\hat J_z+2\ell_I\ell_z^*\hat J_z \hat J_x^n\right)\hat{\mathcal{R}}^\dagger\ket{J,M}\nonumber\\
&\hspace{5cm}=-\frac{N}{2} \left(|\ell_-|^2+|\ell_+|^2+2|\ell_z|^2\right)M^n
-2\Re(\ell_z^*(\ell_+-\ell_-))M^{n+1}.\label{first_diags}
\end{align}
The one-off diagonal elements of the same terms are equal to 
\begin{align}
\bra{J,M-1}& \hat{\mathcal{R}} \left(-\frac12\hat A\hat J_x^n-\frac12 \hat J_x^n\hat A+\sum_{r\in\{+,-\}}\left(\ell_r\ell_I^*\hat J_x^n \hat J_r+\ell_I\ell_r^* \hat J_r^\dagger \hat J_x^n\right)+2\ell_z\ell_I^*\hat J_x^n\hat J_z+2\ell_I\ell_z^*\hat J_z \hat J_x^n\right) \hat{\mathcal{R}}^\dagger\ket{J,M}\nonumber\\
=&\sqrt{(J+M)(J-M+1)}\bigg[M^n\left(-\frac14(\alpha_--\alpha_+-\alpha_z)+\frac12\ell_I\ell_+^*-\frac12\ell_I\ell_-^*-\ell_I\ell_z^*\right)\nonumber\\
&+(M-1)^n\left(-\frac14(\alpha_--\alpha_+-\alpha_z)+\frac12\ell_-\ell_I^*-\frac12\ell_+\ell_I^*-\ell_z\ell_I^*\right)\bigg]\label{first_one_off},
\end{align}
and the $\langle J,M\vert \, \cdot \, \vert J,M+1\rangle$ elements follow similarly. All other elements not on the diagonal nor one-off-diagonals are zero.

We have now accounted for the diagonal and off-diagonal elements in the rotated basis of all but the last term in Eq.~(\ref{jx_evol}). Consider now the last term in Eq.~(\ref{jx_evol}). Since the $\hat L_{j,q}$ operators can be written as a sum over generalized spherical tensor operators [see Eq.~(\ref{spherical_tensor})], one can use the rotational properties of generalized spherical tensors to also write $\hat{\mathcal{R}} \, \hat L_{j,q} \, \hat{\mathcal{R}}^\dagger$ as a sum over generalized spherical tensors. This allows us to derive the rotated version of the last term in Eq.~(\ref{jx_evol}), which reads
\begin{align}
\sum_{j,q,q'}&\ell_q^*\ell_{q'} \hat{\mathcal{R}}\, \hat L^\dagger_{j,q}\, \hat{\mathcal{R}}^\dagger\, \hat J_z^n \, \hat{\mathcal{R}}\, \hat L_{j,q'}\, \hat{\mathcal{R}}^\dagger \nonumber\\
=&\sum_{J,j}\ell_+^*\ell_+\Lambda_{j,1}\Lambda_{j,1}\left(\frac12 \hat T_{1,1}^{J+j,J}-\frac1{\sqrt2} \hat T_{1,0}^{J+j,J}+\frac12 \hat T_{1,-1}^{J+j,J}\right)^\dagger \hat J_z^n\left(\frac12 \hat T_{1,1}^{J+j ,J}-\frac1{\sqrt2} \hat T_{1,0}^{J+j ,J}+\frac12 \hat T_{1,-1}^{J+j ,J}\right) \nonumber\\
&+\ell_+^*\ell_z \Lambda_{j,1}\Lambda_{j ,0}\left(\frac12 \hat T_{1,1}^{J+j,J}-\frac1{\sqrt2} \hat T_{1,0}^{J+j,J}+\frac12 \hat T_{1,-1}^{J+j,J}\right)^\dagger \hat J_z^n\left(\frac1{\sqrt2} \hat T_{1,1}^{J+j ,J}-\frac1{\sqrt2} \hat T_{1,-1}^{J+j ,J}\right) \nonumber\\
&+\ell_+^*\ell_-\Lambda_{j,1}\Lambda_{j,-1}\left(\frac12 \hat T_{1,1}^{J+j,J}-\frac1{\sqrt2} \hat T_{1,0}^{J+j,J}+\frac12 \hat T_{1,-1}^{J+j,J}\right)^\dagger \hat J_z^n\left(\frac12 \hat T_{1,1}^{J+j ,J}+\frac1{\sqrt2} \hat T_{1,0}^{J+j ,J}+\frac12 \hat T_{1,-1}^{J+j ,J}\right) \nonumber\\
&+\ell_z ^*\ell_+\Lambda_{j,0}\Lambda_{j,1}\left(\frac1{\sqrt2} \hat T_{1,1}^{J+j,J}-\frac1{\sqrt{2}} \hat T_{1,-1}^{J+j,J}\right)^\dagger \hat J_z^n\left(\frac12 \hat T_{1,1}^{J+j ,J}-\frac1{\sqrt2} \hat T_{1,0}^{J+j ,J}+\frac12 \hat T_{1,-1}^{J+j ,J}\right) \nonumber\\
&+\ell_z ^*\ell_z \Lambda_{j,0}\Lambda_{j ,0}\left(\frac1{\sqrt2} \hat T_{1,1}^{J+j,J}-\frac1{\sqrt{2}} \hat T_{1,-1}^{J+j,J}\right)^\dagger \hat J_z^n\left(\frac1{\sqrt2} \hat T_{1,1}^{J+j ,J}-\frac1{\sqrt2} \hat T_{1,-1}^{J+j ,J}\right) \nonumber\\
&+\ell_z ^*\ell_-\Lambda_{j,0}\Lambda_{j,-1}\left(\frac1{\sqrt2} \hat T_{1,1}^{J+j,J}-\frac1{\sqrt{2}} \hat T_{1,-1}^{J+j,J}\right)^\dagger \hat J_z^n\left(\frac12 \hat T_{1,1}^{J+j ,J}+\frac1{\sqrt2} \hat T_{1,0}^{J+j ,J}+\frac12 \hat T_{1,-1}^{J+j ,J}\right) \nonumber\\
&+\ell_-^*\ell_+\Lambda_{j,-1}\Lambda_{j,1}\left(\frac12 \hat T_{1,1}^{J+j,J}+\frac1{\sqrt2} \hat T_{1,0}^{J+j,J}+\frac12 \hat T_{1,-1}^{J+j,J}\right)^\dagger \hat J_z^n\left(\frac12 \hat T_{1,1}^{J+j ,J}-\frac1{\sqrt2} \hat T_{1,0}^{J+j ,J}+\frac12 \hat T_{1,-1}^{J+j ,J}\right) \nonumber\\
&+\ell_-^*\ell_z \Lambda_{j,-1}\Lambda_{j ,0}\left(\frac12 \hat T_{1,1}^{J+j,J}+\frac1{\sqrt2} \hat T_{1,0}^{J+j,J}+\frac12 \hat T_{1,-1}^{J+j,J}\right)^\dagger \hat J_z^n\left(\frac1{\sqrt2} \hat T_{1,1}^{J+j ,J}-\frac1{\sqrt2} \hat T_{1,-1}^{J+j ,J}\right) \nonumber\\
&+\ell_-^*\ell_-\Lambda_{j,-1}\Lambda_{j,-1}\left(\frac12 \hat T_{1,1}^{J+j,J}+\frac1{\sqrt2} \hat T_{1,0}^{J+j,J}+\frac12 \hat T_{1,-1}^{J+j,J}\right)^\dagger \hat J_z^n\left(\frac12 \hat T_{1,1}^{J+j ,J}+\frac1{\sqrt2} \hat T_{1,0}^{J+j ,J}+\frac12 \hat T_{1,-1}^{J+j ,J}\right). \label{big_guy}
\end{align}
In the $\ket{J,M_z}$ basis, the above matrix has diagonal terms and one-off diagonal terms. At first glance it looks like there will also be non-vanishing two-off diagonal terms (in the form of $\mel{J,M-2}{\;\;\cdot\;\;}{J,M}$ and $\mel{J,M}{\;\;\cdot\;\;}{J,M-2}$). However, these two-off diagonal elements cancel due to the sum over values of $j$, hence we only need to concern ourselves with the diagonal and one-off diagonal terms.

The diagonal elements of Eq.~(\ref{big_guy}) are
\begin{align}
\bra{J,M}&\sum_{j,q,q'}\ell_q^*\ell_{q'} \hat{\mathcal{R}}\, \hat L^\dagger_{j,q}\, \hat{\mathcal{R}}^\dagger\, \hat J_z^n\, \hat{\mathcal{R}} \, \hat L_{j,q'}\, \hat{\mathcal{R}}^\dagger\ket{J,M}\nonumber\\
=&\frac{1}{8} \bigg(( M -1)^n (2  M + N ) ( 2\ell_z - \ell_- + \ell_+ ) \left( 2\ell_z ^*- \ell_- ^*+ \ell_+ ^*\right) \nonumber\\
&-( M +1)^n (2  M - N ) ( 2\ell_z + \ell_- - \ell_+ ) \left( 2\ell_z ^*+ \ell_- ^*- \ell_+ ^*\right) + 2 M^n N ( \ell_- + \ell_+ )  \left( \ell_- ^*+ \ell_+ ^*\right)\bigg). \label{diagonal_terms}
\end{align}
Grouping by powers of $M$ and using the binomial theorem, we find that Eq.~(\ref{diagonal_terms}) can be further simplified into
\begin{align}
\bra{J,M}&\sum_{j,q,q'}\ell_q^*\ell_{q'} \hat{\mathcal{R}}\, \hat L^\dagger_{j,q}\, \hat{\mathcal{R}}^\dagger\, \hat J_z^n\, \hat{\mathcal{R}} \, \hat L_{j,q'}\, \hat{\mathcal{R}}^\dagger\ket{J,M}\nonumber\\
&=\frac{1}{8} \bigg(( M -1)^n (2  M + N ) Q -( M +1)^n (2  M - N ) Y+2  M ^n N  S\bigg) \nonumber\\
&=\frac{1}{8}\sum_{k=0}^n M^k\binom{n}{k}\left[(-1)^{n-k}(2M+N)Q-(2M-N)Y\right]+\frac14 M^n NS,
\end{align}
where we have defined the coefficients $Q=|2\ell_z - \ell_- + \ell_+|^2$, $Y=|2\ell_z + \ell_- - \ell_+|^2$, and $S= |\ell_- + \ell_+|^2$. This expression can be further simplified to read
\begin{align}
\bra{J,M}&\sum_{j,q,q'}\ell_q^*\ell_{q'} \hat{\mathcal{R}}\, \hat L^\dagger_{j,q}\, \hat{\mathcal{R}}^\dagger\, \hat J_z^n\, \hat{\mathcal{R}} \, \hat L_{j,q'}\, \hat{\mathcal{R}}^\dagger\ket{J,M}\nonumber\\
&=\frac18\left[2\sum_{k=0}^n M^{k+1}\binom{n}{k}\left(Q(-1)^{n-k}-Y\right)+N\sum_{k=0}^n M^k\binom{n}{k}\left(Q(-1)^{n-k}+Y\right)\right]+\frac14 M^n NS \nonumber\\
&=\frac18\left[2\sum_{k=0}^n M^{k}\binom{n}{k-1}\left(-Q(-1)^{n-k}-Y\right)+N\sum_{k=0}^n M^k\binom{n}{k}\left(Q(-1)^{n-k}+Y\right)\right] \nonumber\\
&+\frac14 M^n NS+2\Re(\ell_z^*(\ell_+-\ell_-))M^{n+1} \nonumber\\
&=\frac{1}{8}\sum_{k=0}^nM^k\left[\binom{n}{k}\left(NQ(-1)^{n-k}+NY\right)-\binom{n}{k-1}\left(2Q(-1)^{n-k}+2Y\right)\right]+\frac14NSM^n\nonumber\\
&+2\Re(\ell_z^*(\ell_+-\ell_-))M^{n+1}.\label{second_diags}
\end{align}
We can therefore write that
\begin{align}
\sum_{j,q,q'}\ell_q^*\ell_{q'}\hat R_y\hat L^\dagger_{j,q}\hat R_y^\dagger \hat J_z^n \hat R_y\hat L_{j,q'}\hat R_y^\dagger=\sum_kc_k\hat J_z^n+\hat\theta
\end{align}
with
\begin{align}
c_k=\frac18\left[\binom{n}{k}\left(NQ(-1)^{n-k}-NY\right)+\binom{n}{k-1}\left(-2Q(-1)^{n-k}+2Y\right)\right]+\frac14NS\delta_{n,k}+2\Re(\ell_z^*(\ell_+-\ell_-))\delta_{k,n+1},
\end{align}
and where $\hat\theta$ represents the non-diagonal terms (specifically those leading to one-off diagonal elements) which we will derive below. We shall see,  however, that $\hat \theta$ vanishes in many special cases, including the case of optical pumping studied in the main text.

Let us briefly consider the sum of Eqs.~(\ref{first_diags}) and (\ref{second_diags}), which yields us all diagonal elements $\hat{\mathcal{R}}\, e^{\mathcal L^\dagger t} \hat J_x^n\, \hat{\mathcal{R}}^\dagger$. Here $\mathcal L^\dagger$ is the adjoint master equation superoperator from Eq.~(\ref{jx_evol}). We find that the terms of order $M^{n+1}$ cancel, and truncating to only the three largest order of $M$ terms the sum of Eqs.~(\ref{first_diags}) and (\ref{second_diags}) is
\begin{align}
    \bra{J,M}\hat{\mathcal{R}}\, & e^{\mathcal L^\dagger t} \hat J_x^n \, \hat{\mathcal{R}}^\dagger\ket{J,M}\nonumber\\
    &= -\frac12 n\omega M^n-n(N-n+1)\Re(\ell_z^*(\ell_+-\ell_-))M^{n-1}+\frac1{24}n(n-1)(3N-2n+4)\omega M^{n-2} \nonumber\\
    &\approx -\frac12 n\omega M^n-nN\Re(\ell_z^*(\ell_+-\ell_-))M^{n-1}+\frac18 n(n-1)N\omega M^{n-2},\label{diagonal_sum}
\end{align}
where in the second line we have made a large $N$ approximation (i.e., $N\gg n$), and defined $\omega=4|\ell_z|^2+|\ell_+-\ell_-|^2$. The form of Eq.~(\ref{diagonal_sum}) allows us to write the rotated time-evolved $\hat J_x^n$-operator as
\begin{equation}
    \hat{\mathcal{R}} e^{\mathcal L^\dagger t} \hat J_z^n \hat{\mathcal{R}}^\dagger = -\frac12 n\omega \hat J_z^n-nN\Re(\ell_z^*(\ell_+-\ell_-))\hat J_z^{n-1}+\frac18 n(n-1)N\omega \hat J_z^{n-2}+(\text{off-diagonal terms}).\label{all_but_off_diagonals_zaxis}
\end{equation}
Applying the inverse rotations to Eq.~(\ref{all_but_off_diagonals_zaxis}) to rotate back into the ``$\hat J_x$''-direction we find
\begin{align}
    e^{\mathcal L^\dagger t}\hat J_x^n =& \hat{\mathcal{R}}^\dagger \left(\hat{\mathcal{R}} e^{\mathcal L^\dagger t} \hat J_z^n \hat{\mathcal{R}}^\dagger \right) \hat{\mathcal{R}} \nonumber\\
    =& -\frac12 n\omega \hat J_x^n-nN\Re(\ell_z^*(\ell_+-\ell_-))\hat J_x^{n-1}+\frac18 n(n-1)N\omega \hat J_x^{n-2}+ \hat{\mathcal{R}}^\dagger\left(\text{off-diagonal terms}\right)\hat{\mathcal{R}},\label{all_but_off_diagonals}
\end{align}
where in this case ``off-diagonal terms'' refers to terms in the rotated frame. Note that since $\expv{\hat J_x^n}\sim N^{n/2}$, the first and third terms on the RHS of Eq.~(\ref{all_but_off_diagonals}) are on the order of $N^{n/2}$, whereas the second term is on the order of $N^{n/2+1/2}$. This means that for systems where $\Re(\ell_z^*(\ell_+-\ell_-))$ is nonzero, this term will tend to dominate the behavior for later times.

Next we consider the off-diagonal terms, some of which have already been calculated in Eq.~(\ref{first_one_off}). For Eq.~(\ref{big_guy}) the one-off diagonal elements (which are contained in the matrix $\hat\theta$ as $\mel{J,M-1}{\hat\theta}{J,M}$) are given by
\begin{align}
\bra{J,M-1}&\sum_{j,q,q'}\ell_q^*\ell_{q'} \hat{\mathcal{R}} \hat L^\dagger_{j,q} \hat{\mathcal{R}}^\dagger \hat J_z^n \hat{\mathcal{R}} \hat L_{j,q'} \hat{\mathcal{R}}^\dagger\ket{J,M}\nonumber\\
&=\frac{1}{4} \sqrt{(J-M+1) (J+M)} \Big[(M-1)^n ( 2\ell_z - \ell_- + \ell_+ ) \left( \ell_- ^*+ \ell_+ ^*\right)- M^n\left( 2\ell_z ^*+ \ell_- ^*- \ell_+ ^*\right)( \ell_- + \ell_+ ) \Big].\label{second_one_off}
\end{align}
We can now sum all the one-off diagonal elements from Eqs.~(\ref{first_one_off}) and (\ref{second_one_off}) to find the one-off diagonal elements for $\hat{\mathcal{R}} e^{t\mathcal L^\dagger}J_x^n \hat{\mathcal{R}}^\dagger$. We find that 
\begin{align}
\bra{J,M-1}\hat{\mathcal{R}} &e^{\mathcal L^\dagger t} \hat J_x^n \hat{\mathcal{R}}^\dagger \ket{J,M} \nonumber\\
=&\frac{1}{4} \sqrt{(J-M+1) (J+M)} \left((M-1)^n-M^n\right)\nonumber\\
&\times\left( \ell_z ^* (2  \ell_I + \ell_- + \ell_+ )+ \ell_I ^* (-2  \ell_z + \ell_- - \ell_+ )+ \ell_- ^* ( \ell_z + \ell_I + \ell_+ )+ \ell_+ ^* ( \ell_z - \ell_I - \ell_- )\right) \nonumber\\
=&\frac{1}{4} \sqrt{(J-M+1) (J+M)}\left((M-1)^n- M^n)\right)\phi,
\end{align}
where we have defined $\phi=\left( \ell_z ^* (2  \ell_I + \ell_- + \ell_+ )+ \ell_I ^* (-2  \ell_z + \ell_- - \ell_+ )+ \ell_- ^* ( \ell_z + \ell_I + \ell_+ )+ \ell_+ ^* ( \ell_z - \ell_I - \ell_- )\right)$. We can therefore write the operator $\hat\Theta$ containing all one-off diagonal elements of $\hat{\mathcal{R}} e^{t\mathcal L^\dagger} \hat J_x^n \hat{\mathcal{R}}^\dagger$ as
\begin{align}
\hat\Theta=&\frac14\sum_{J,M}\sqrt{(J-M+1)(J+M)}\left[(M-1)^n-M^n\right]\phi\ketbra{J,M-1}{J,M} \nonumber\\
&+\sqrt{(J-M)(J+M+1)}\left[M^n-(M+1)^n\right]\phi^*\ketbra{J,M+1}{J,M},
\end{align}
which can be simplified using the binomial theorem to
\begin{equation}
\hat\Theta=\sum_{k=0}^n \underbrace{\frac14\left[(\phi^*\hat J_+-\phi \hat J_-)\delta_{n,k}-\binom{n}{k}\left(\phi^*\hat J_+-(-1)^{n-k}\phi \hat J_-\right)\right]}_{=: \hat D_k} \hat J_z^k = \sum_{k=0}^n\hat D_k\hat J_z^k.\label{Dk_def}
\end{equation}
The $\hat D_k$ operator, defined implicitly by Eq.~(\ref{Dk_def}), has a few noteworthy properties. The first is that $\hat D_n=0$, which can be seen by inspection of Eq.~(\ref{Dk_def}). For $k < n$ we find that
\begin{equation}
\hat D_k=-\frac14\binom{n}{k}\left(\phi^*\hat J_+-(-1)^{n-k}\phi \hat J_-\right).
\end{equation}
This shows that $\hat D_k$ is simply a rotated and rescaled $\hat J_x$ operator in the $\hat J_x$-$\hat J_y$ plane, rotated by an angle determined by the phase of $\phi$. We can therefore rewrite $\hat D_k$ as
\begin{align}
\label{nk_odd}\hat D_{k}^{\text{ ($n-k$ odd)}}&=-\frac12\binom{n}{k}\left(\Re(\phi)\hat J_x+\Im(\phi)\hat J_y\right),\\
\label{nk_even}\hat D_{k}^{\text{ ($n-k$ even)}}&=-\frac i2\binom{n}{k}\left(-\Im(\phi)\hat J_x+\Re(\phi)\hat J_y\right).
\end{align}

We now apply the inverse rotation $\hat{\mathcal{R}}^\dagger$ to Eq.~(\ref{Dk_def}) to rotate back into the ``$\hat J_x$ direction''. We further make the large $N$ approximation (i.e. $N \gg n$) and drop any terms of lower order than $N^{n/2}$, the largest order in $N$. The $k=n-1$ and $k=n-2$ terms contain this order of $N$, hence after our truncation we find that
\begin{equation}
\bigg\langle\sum_k \hat{\mathcal{R}}^\dagger \hat D_k \hat{\mathcal{R}} \hat J_x^k\bigg\rangle
\approx\frac n2\Re(\phi)\expv{\hat J_z\hat J_x^{n-1}}-\frac n2\Im(\phi)\expv{\hat J_y\hat J_x^{n-1}}-\frac i4n(n-1)\Im(\phi)\expv{\hat J_z\hat J_x^{n-2}}-\frac i4n(n-1)\Re(\phi)\expv{\hat J_y\hat J_x^{n-2}}.\label{dk_analysis}
\end{equation}
\end{widetext}
For large $N$ we can approximate $\expv{\hat J_z\hat J_x^k}\approx\expv{\hat J_z}\expv{\hat J_x^k}$. Additionally, the last term in Eq.~(\ref{dk_analysis}) is on the order of $N^{n/2-1}$, so we can approximate it to be zero for sufficiently large $N$. Next, note that the quantity in Eq.~(\ref{dk_analysis}) must be real, implying that the sum over imaginary part terms vanishes. Although we are truncating the sum over $k$ values, we can still account for this by simply taking the real part of all terms in Eq.~(\ref{dk_analysis}). Since the second to last term can be approximated as the product of $\expv{\hat J_z}$ and $\expv{\hat J_x^{n-2}}$, which are both real, we see that this term is purely imaginary and thus vanishes.

Equation~(\ref{dk_analysis}) may therefore be approximated as
\begin{align}
\bigg\langle\sum_k \hat{\mathcal{R}}^\dagger \hat D_k \hat{\mathcal{R}} \hat J_x^k\bigg\rangle \approx&
\frac n2\Re(\phi)\expv{\hat J_z}\expv{\hat J_x^{n-1}}\nonumber\\
&-\frac n2\Im(\phi)\Re\left[\expv{\hat J_y\hat J_x^{n-1}}\right].\label{almost_done}
\end{align}
Finally, we are able to make one last approximation which greatly simplifies the above expression. In the Holstein-Primakoff approximation one finds (by deriving the Weyl symbol of $\hat P\hat X^{n-1}$) that
\begin{align}
    \Re[\langle\hat J_y\hat J_x^{n-1}\rangle]\approx\langle\hat J_y\rangle\langle\hat J_x^{n-1}\rangle,\label{eq:hp_appx_of_expectation}
\end{align}
and therefore Eq.~(\ref{almost_done}) can be simplified to
\begin{align}
    \bigg\langle\sum_k \hat{\mathcal{R}}^\dagger \hat D_k &\hat{\mathcal{R}} \hat J_x^k\bigg\rangle \nonumber\\
    \approx& \frac{n}{2}\expv{\hat J_x^{n-1}}\expval{\Re(\phi)\hat J_z-\Im(\phi)\hat J_y} \nonumber\\
    =&|\phi|\frac n2\expv{\hat J_x^{n-1}}\expval{\cos(\alpha)\hat J_z-\sin(\alpha)\hat J_y} \nonumber\\
    =&|\phi|\frac n2\expv{\hat J_x^{n-1}}\expval{\hat R_x(\alpha)\hat J_z\hat R_x^\dagger(\alpha)},
\end{align}
where $\phi=|\phi|e^{i\alpha}$ and $\hat R_x(\alpha)=\exp[-i\alpha\hat J_x]$.

Combining this with our calculation for $\expv{\hat J_x^n}$'s dependence on the diagonal terms we find that
\begin{align}
\frac{\mathrm d}{\mathrm dt}\expv{\hat J_x^n}\approx&-\frac12n\omega \expv{\hat J_x^n}+\frac N8n(n-1)\omega \expv{\hat J_x^{n-2}}\nonumber\\
&-nN\Re(\ell_z^*(\ell_+-\ell_-))\expv{\hat J_x^{n-1}}\nonumber\\
&+|\phi|\frac n2\expv{\hat J_x^{n-1}}\expval{\hat R_x(\alpha)\hat J_z\hat R_x^\dagger(\alpha)}.\label{whole_fp}
\end{align}
The first two terms in the above expression are exactly what we obtained in the case of pure optical pumping, whereas the last two terms come from generalizing to an arbitrary local operator $\boldsymbol\ell^{(i)}$. While it may not be immediately obvious, Eq.~(\ref{whole_fp}) is a Fokker-Planck equation. This can be understood by noting that $\expv{\hat R_x\hat J_z\hat R_x^\dagger}$ in Eq.~(\ref{whole_fp}) is some function of time which can be found by deriving the equations of motion of the mean spin axes $\langle\hat J_x\rangle(t)$, $\langle\hat J_y\rangle(t)$, and $\langle\hat J_z\rangle(t)$. These functions can be found by solving a set of 3 coupled differential equations, which are derived in App.~\ref{evol_jy}. Therefore Eq.~(\ref{whole_fp}) is of the form
\begin{align}
    \frac{\mathrm d}{\mathrm dt}\langle\hat J_x^n\rangle=&n(n-1)D\langle\hat J_x^{n-2}\rangle+n\mu_1\langle\hat J_x^n\rangle \nonumber\\
    &+n\mu_2(t)\langle\hat J_x^{n-1}\rangle,
\end{align}
where $D$ is the diffusion coefficient and $\mu(x,t)=x \mu_1+\mu_2(t)$ is the drift coefficient. See App.~\ref{det_fp} for more details on the form of this particular equation of motion.

\begin{figure}
    \centering
    \includegraphics[width=0.95\linewidth]{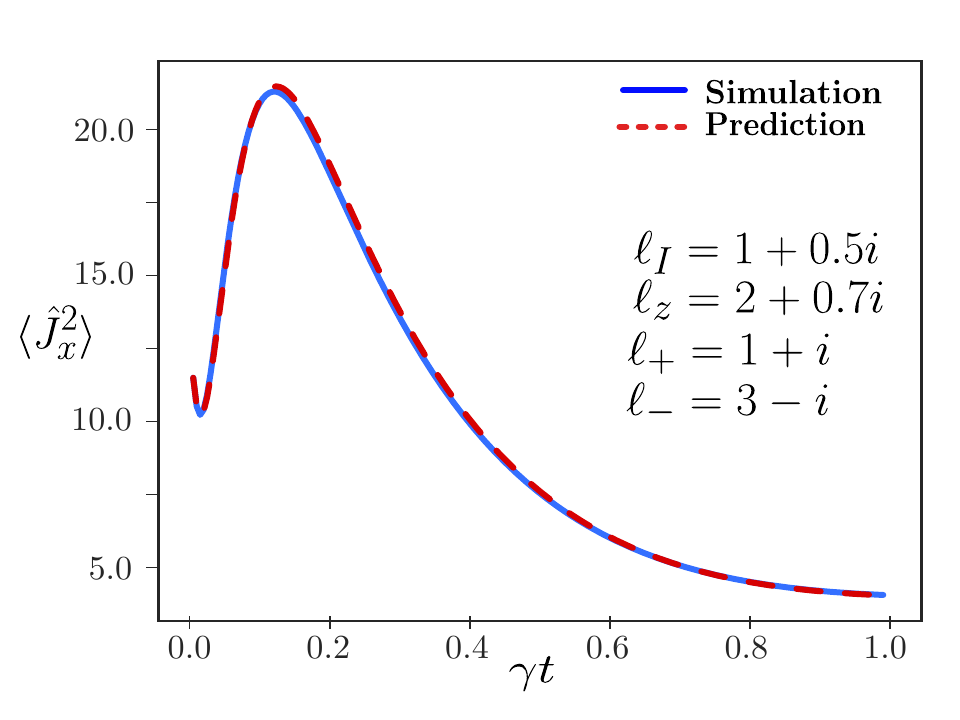}
    \caption{Plot of $\expv{\hat J_x^2}$ for an initial first excited Dicke state $\ketbra{J=8,M=7}$ of $N=16$ spins. The blue line is found by directly simulating the master equation [Eq.~(\ref{master2})], and the red dashed line is the prediction made by evolving $\expv{\hat J_x^2}$ under Eq.~(\ref{whole_fp}). The state evolves under the randomly chosen local operator defined by $\ell_I=1+0.5i$, $\ell_z=2+0.7i$, $\ell_+=1+i$, and $\ell_-=3-i$. These coefficients were chosen to demonstrate that even when $\phi\neq0$ and $\Re(\ell_z^*(\ell_+-\ell_-))\neq0$ the evolution of $\expv{\hat J_x^n}$ tends to be well described by Eq.~(\ref{whole_fp}) for long times on the order of $\gamma t\sim1$.} 
    \label{fig:arbitrary_local_decoherence}
\end{figure}

In Fig.~\ref{fig:arbitrary_local_decoherence} we plot the evolution of the second moment of $\hat J_x$ for an initial Dicke state with $N=16$ as predicted by Eq.~(\ref{whole_fp}). We compare this to the value of $\expv{\hat J_x^2}$ obtained by simply simulating the master equation [Eq.~(\ref{master2})]. We find that even for moderately sized systems Eq.~(\ref{whole_fp}) tends to be a good approximation, even for timescales on the order of $\gamma t\sim1$.

Finally, while Eq.~(\ref{eq:hp_appx_of_expectation}) assumes the Holstein-Primakoff approximation, we have found that numerical simulations suggest that even outside of the Holstein-Primakoff regime Eq.~(\ref{whole_fp}) tends to converge for moderate $N\sim100$. We believe this is due to the system becoming more Gaussian with time, and thus observables tend to become less correlated.

\section{Time evolution of arbitrary spin operators}
\label{evol_jy}
In Sec.~\ref{subsec:genlocaldecoh} we studied the extension of our analytical results to general local decoherence, and in App.~\ref{generalization} we derived the equation of motion for arbitrary moments $\langle \hat J_x^n \rangle$ of the $\hat J_x$-operator. In this Appendix we will show one particular method of deriving the equations of motion for moments of $\hat J_y$ and $\hat J_z$ using the already derived results for $\hat J_x$ in App.~\ref{generalization}.

In Eq.~(\ref{whole_fp}) of App.~\ref{generalization} we obtained the equation of motion for the $n$th moment $\langle \hat J_x^n \rangle$ by considering the evolution under the generalized decoherence specified by the local operator $\hat \ell = \sum_i \hat \ell^{(i)} = \sum_i \ell_I\mathds1^{(i)}+\ell_+\hat\sigma_+^{(i)}+\ell_-\hat\sigma_-^{(i)}+\ell_z\hat\sigma_z^{(i)}$ defined in Eq.~(\ref{app:eq:generaldecohoper}). Since $\hat \ell$ can be chosen arbitrarily, our strategy for deriving the evolution of the moments of $\hat J_y$ and $\hat J_z$ will be to use rotations to transform our coordinates such that we may reuse the derivation in App.~\ref{generalization}.

To obtain the time evolution of $\hat J_\mu^n$ we perform a passive rotation on the local jump operators $\hat{\ell}^{(i)}$. For example, if one wishes to obtain the evolution of the $\hat J_x^n$ operator rotated by the polar angle $\theta$ and the azimuthal angle $\varphi$, $\hat J_{\theta,\varphi}^n = \hat R_y(\varphi)\hat R_z(\theta)\hat J_x^n\hat R_z^\dagger(\theta)\hat R_y^\dagger(\varphi)$, one must consider the passively rotated local operator $\hat{\ell}^{(i)}_{\theta,\varphi} = \hat R_y^\dagger(\varphi)\hat R_z^\dagger(\theta)\hat{\ell}^{(i)}\hat R_z(\theta)\hat R_y(\varphi)$. The evolution of $\hat J_{\theta,\varphi}^n$ with respect to $\ell^{(i)}_{\theta,\varphi}$ is then identical to that of $\hat J_x^n$ with respect to $\ell^{(i)}$. It will be sufficient to derive the equations of motion for $\hat J_y^n$ and $\hat J_z^n$, which is what we will focus on in this section.

Consider first $\hat J_y^n$. We note that the rotation $\hat R_z(-\pi/2)$ takes $\hat J_x\to \hat J_y$. Applying the inverse rotation to $\hat{\ell}^{(i)}$ we find that 
\begin{align}
    \hat R_z(\pi/2)\hat\sigma_+\hat R_z^\dagger(\pi/2)&=i\hat \sigma_+,\\
    \hat R_z(\pi/2)\hat\sigma_-\hat R_z^\dagger(\pi/2)&=-i\hat \sigma_-.
\end{align}
Therefore we can represent the application of $\hat R_z(\pi/2)$ on $\hat{\ell}^{(i)}$ by the replacement rules
\begin{align}
    \ell_+&\to -i\ell_+,\\
    \ell_-&\to i\ell_-.
\end{align}
Applying these rules to the coefficients in Eq.~(\ref{whole_fp}) we find the mapping of the parameters
\begin{align}
    &\omega\to \omega_y = 4|\ell_z|^2+|\ell_++\ell_-|^2, \\
    &\phi\to  \phi_y = \ell_z ^* (2  \ell_I + i\ell_- - i\ell_+ )+ \ell_I ^* (-2  \ell_z + i\ell_- + i\ell_+ )\nonumber\\
    &\hspace{1.5cm}+ i\ell_- ^* ( \ell_z + \ell_I - i\ell_+ )- i\ell_+ ^* ( \ell_z - \ell_I - i\ell_- ),
\end{align}
and likewise the following mapping for the coefficient of the third term in Eq.~(\ref{whole_fp})
\begin{equation}
    \Re(\ell_z^*(\ell_+-\ell_-))\to i\Im(\ell_z^*(\ell_++\ell_-)).
\end{equation}
Therefore the equation of motion for $\expv{\hat J_y^n}$ is
\begin{align}
    \frac{\mathrm d}{\mathrm dt}\expv{\hat J_y^n}\approx&-\frac12n\omega_y \expv{\hat J_y^n}+\frac N8n(n-1)\omega_y \expv{\hat J_y^{n-2}}\nonumber\\
    &-nN\Im(\ell_z^*(\ell_++\ell_-))\expv{\hat J_y^{n-1}}\nonumber\\
    &+\frac{n}{2}\expv{\hat J_y^{n-1}}\Re(\phi_y)\expv{\hat J_z}\nonumber\\
    &+\frac{n}{2}\expv{\hat J_y^{n-1}}\Im(\phi_y)\expv{\hat J_x}\label{jy_evol}.
\end{align}
Note that above we have also relabeled $\hat J_x\to\hat J_y$ and $\hat J_y\to-\hat J_x$.

Similarly we find the equation of motion for $\hat J_z^n$, by using the rotation $\hat R_y(-\pi/2)$, which takes $\hat J_x\to\hat J_z$. We therefore apply the inverse rotation to $\hat{\ell}^{(i)}$, and find that
\begin{align}
    \hat R_y(\pi/2)\hat \sigma_+\hat R_y^\dagger(\pi/2)&=-\frac12\hat\sigma_z+\frac12\hat\sigma_+-\frac12\hat\sigma_-,\\
    \hat R_y(\pi/2)\hat \sigma_-\hat R_y^\dagger(\pi/2)&=-\frac12\hat\sigma_z-\frac12\hat\sigma_++\frac12\hat\sigma_-,\\
    \hat R_y(\pi/2)\hat \sigma_z\hat R_y^\dagger(\pi/2)&=\hat\sigma_++\hat\sigma_-.
\end{align}
Once again we can represent this transformation on $\hat{\ell}^{(i)}$ using a set of replacement rules
\begin{align}
    \ell_+&\to\frac12\ell_+-\frac12\ell_-+\ell_z,\\
    \ell_-&\to\frac12\ell_--\frac12\ell_++\ell_z,\\
    \ell_z&\to-\frac12\ell_+-\frac12\ell_-.
\end{align}
Applying these replacement rules to the coefficients in Eq.~(\ref{whole_fp}) we find that the parameters $\omega$ and $\phi$ get mapped to
\begin{align}
    &\omega\to\omega_z=2|\ell_+|^2+2|\ell_-|^2,\\
    &\phi\to\phi_z=2[\ell_-(\ell_I^*-\ell_z^*)-\ell_+^*(\ell_I+\ell_z)],
\end{align}
and the third term in Eq.~(\ref{whole_fp}) gets mapped as
\begin{equation}
    \Re(\ell_z^*(\ell_+-\ell_-))\to\frac12|\ell_-|^2-\frac12|\ell_+|^2.
\end{equation}
The equation of motion for $\langle\hat J_z^n\rangle$ is therefore
\begin{align}
    \frac{\mathrm d}{\mathrm dt}\langle\hat J_z^n\rangle=&-\frac12n\omega_z \expv{\hat J_z^n}+\frac N8n(n-1)\omega_z \expv{\hat J_z^{n-2}}\nonumber\\
    &-nN\Re(\ell_z^*(\ell_+-\ell_-))\expv{\hat J_z^{n-1}}\nonumber\\
    &-\frac{n}{2}\expv{\hat J_z^{n-1}}\Re(\phi_z)\expv{\hat J_x}\nonumber\\
    &-\frac{n}{2}\expv{\hat J_z^{n-1}}\Im(\phi_z)\expv{\hat J_y},\label{jz_evol}
\end{align}
where again we have relabeled $\hat J_x\to\hat J_z$ and $\hat J_z\to-\hat J_x$.

\bibliography{refs.bib}

\begin{thebibliography}{27}%
\makeatletter
\providecommand \@ifxundefined [1]{%
 \@ifx{#1\undefined}
}%
\providecommand \@ifnum [1]{%
 \ifnum #1\expandafter \@firstoftwo
 \else \expandafter \@secondoftwo
 \fi
}%
\providecommand \@ifx [1]{%
 \ifx #1\expandafter \@firstoftwo
 \else \expandafter \@secondoftwo
 \fi
}%
\providecommand \natexlab [1]{#1}%
\providecommand \enquote  [1]{``#1''}%
\providecommand \bibnamefont  [1]{#1}%
\providecommand \bibfnamefont [1]{#1}%
\providecommand \citenamefont [1]{#1}%
\providecommand \href@noop [0]{\@secondoftwo}%
\providecommand \href [0]{\begingroup \@sanitize@url \@href}%
\providecommand \@href[1]{\@@startlink{#1}\@@href}%
\providecommand \@@href[1]{\endgroup#1\@@endlink}%
\providecommand \@sanitize@url [0]{\catcode `\\12\catcode `\$12\catcode
  `\&12\catcode `\#12\catcode `\^12\catcode `\_12\catcode `\%12\relax}%
\providecommand \@@startlink[1]{}%
\providecommand \@@endlink[0]{}%
\providecommand \url  [0]{\begingroup\@sanitize@url \@url }%
\providecommand \@url [1]{\endgroup\@href {#1}{\urlprefix }}%
\providecommand \urlprefix  [0]{URL }%
\providecommand \Eprint [0]{\href }%
\providecommand \doibase [0]{http://dx.doi.org/}%
\providecommand \selectlanguage [0]{\@gobble}%
\providecommand \bibinfo  [0]{\@secondoftwo}%
\providecommand \bibfield  [0]{\@secondoftwo}%
\providecommand \translation [1]{[#1]}%
\providecommand \BibitemOpen [0]{}%
\providecommand \bibitemStop [0]{}%
\providecommand \bibitemNoStop [0]{.\EOS\space}%
\providecommand \EOS [0]{\spacefactor3000\relax}%
\providecommand \BibitemShut  [1]{\csname bibitem#1\endcsname}%
\let\auto@bib@innerbib\@empty
\bibitem [{\citenamefont {Chaneli\'{e}re}\ \emph {et~al.}(2018)\citenamefont
  {Chaneli\'{e}re}, \citenamefont {H\'{e}tet},\ and\ \citenamefont
  {Sangouard}}]{atomic_memory2018}%
  \BibitemOpen
  \bibfield  {author} {\bibinfo {author} {\bibfnamefont {T.}~\bibnamefont
  {Chaneli\'{e}re}}, \bibinfo {author} {\bibfnamefont {G.}~\bibnamefont
  {H\'{e}tet}}, \ and\ \bibinfo {author} {\bibfnamefont {N.}~\bibnamefont
  {Sangouard}},\ }\href {\doibase https://doi.org/10.1016/bs.aamop.2018.02.002}
  {\bibfield  {journal} {\bibinfo  {journal} {Advances In Atomic, Molecular,
  and Optical Physics}\ }\textbf {\bibinfo {volume} {67}},\ \bibinfo {pages}
  {77} (\bibinfo {year} {2018})}\BibitemShut {NoStop}%
\bibitem [{\citenamefont {Pezz\`e}\ \emph {et~al.}(2018)\citenamefont
  {Pezz\`e}, \citenamefont {Smerzi}, \citenamefont {Oberthaler}, \citenamefont
  {Schmied},\ and\ \citenamefont {Treutlein}}]{Pezze2018}%
  \BibitemOpen
  \bibfield  {author} {\bibinfo {author} {\bibfnamefont {L.}~\bibnamefont
  {Pezz\`e}}, \bibinfo {author} {\bibfnamefont {A.}~\bibnamefont {Smerzi}},
  \bibinfo {author} {\bibfnamefont {M.~K.}\ \bibnamefont {Oberthaler}},
  \bibinfo {author} {\bibfnamefont {R.}~\bibnamefont {Schmied}}, \ and\
  \bibinfo {author} {\bibfnamefont {P.}~\bibnamefont {Treutlein}},\ }\href
  {\doibase 10.1103/RevModPhys.90.035005} {\bibfield  {journal} {\bibinfo
  {journal} {Rev. Mod. Phys.}\ }\textbf {\bibinfo {volume} {90}},\ \bibinfo
  {pages} {035005} (\bibinfo {year} {2018})}\BibitemShut {NoStop}%
\bibitem [{\citenamefont {Borish}\ \emph {et~al.}(2020)\citenamefont {Borish},
  \citenamefont {Markovi\ifmmode~\acute{c}\else \'{c}\fi{}}, \citenamefont
  {Hines}, \citenamefont {Rajagopal},\ and\ \citenamefont
  {Schleier-Smith}}]{Schleier-Smith2020}%
  \BibitemOpen
  \bibfield  {author} {\bibinfo {author} {\bibfnamefont {V.}~\bibnamefont
  {Borish}}, \bibinfo {author} {\bibfnamefont {O.}~\bibnamefont
  {Markovi\ifmmode~\acute{c}\else \'{c}\fi{}}}, \bibinfo {author}
  {\bibfnamefont {J.~A.}\ \bibnamefont {Hines}}, \bibinfo {author}
  {\bibfnamefont {S.~V.}\ \bibnamefont {Rajagopal}}, \ and\ \bibinfo {author}
  {\bibfnamefont {M.}~\bibnamefont {Schleier-Smith}},\ }\href {\doibase
  10.1103/PhysRevLett.124.063601} {\bibfield  {journal} {\bibinfo  {journal}
  {Phys. Rev. Lett.}\ }\textbf {\bibinfo {volume} {124}},\ \bibinfo {pages}
  {063601} (\bibinfo {year} {2020})}\BibitemShut {NoStop}%
\bibitem [{\citenamefont {Wesenberg}\ \emph {et~al.}(2009)\citenamefont
  {Wesenberg}, \citenamefont {Ardavan}, \citenamefont {Briggs}, \citenamefont
  {Morton}, \citenamefont {Schoelkopf}, \citenamefont {Schuster},\ and\
  \citenamefont {M{\o}lmer}}]{computing1}%
  \BibitemOpen
  \bibfield  {author} {\bibinfo {author} {\bibfnamefont {J.~H.}\ \bibnamefont
  {Wesenberg}}, \bibinfo {author} {\bibfnamefont {A.}~\bibnamefont {Ardavan}},
  \bibinfo {author} {\bibfnamefont {G.~A.~D.}\ \bibnamefont {Briggs}}, \bibinfo
  {author} {\bibfnamefont {J.~J.~L.}\ \bibnamefont {Morton}}, \bibinfo {author}
  {\bibfnamefont {R.~J.}\ \bibnamefont {Schoelkopf}}, \bibinfo {author}
  {\bibfnamefont {D.~I.}\ \bibnamefont {Schuster}}, \ and\ \bibinfo {author}
  {\bibfnamefont {K.}~\bibnamefont {M{\o}lmer}},\ }\href {\doibase
  10.1103/physrevlett.103.070502} {\bibfield  {journal} {\bibinfo  {journal}
  {Physical Review Letters}\ }\textbf {\bibinfo {volume} {103}},\ \bibinfo
  {pages} {070502} (\bibinfo {year} {2009})}\BibitemShut {NoStop}%
\bibitem [{\citenamefont {Mu\~noz Arias}\ \emph {et~al.}(2020)\citenamefont
  {Mu\~noz Arias}, \citenamefont {Poggi}, \citenamefont {Jessen},\ and\
  \citenamefont {Deutsch}}]{Munoz2010}%
  \BibitemOpen
  \bibfield  {author} {\bibinfo {author} {\bibfnamefont {M.~H.}\ \bibnamefont
  {Mu\~noz Arias}}, \bibinfo {author} {\bibfnamefont {P.~M.}\ \bibnamefont
  {Poggi}}, \bibinfo {author} {\bibfnamefont {P.~S.}\ \bibnamefont {Jessen}}, \
  and\ \bibinfo {author} {\bibfnamefont {I.~H.}\ \bibnamefont {Deutsch}},\
  }\href {\doibase 10.1103/PhysRevLett.124.110503} {\bibfield  {journal}
  {\bibinfo  {journal} {Phys. Rev. Lett.}\ }\textbf {\bibinfo {volume} {124}},\
  \bibinfo {pages} {110503} (\bibinfo {year} {2020})}\BibitemShut {NoStop}%
\bibitem [{\citenamefont {Leroux}\ \emph {et~al.}(2010)\citenamefont {Leroux},
  \citenamefont {Schleier-Smith},\ and\ \citenamefont
  {Vuleti\ifmmode~\acute{c}\else \'{c}\fi{}}}]{Leroux2010}%
  \BibitemOpen
  \bibfield  {author} {\bibinfo {author} {\bibfnamefont {I.~D.}\ \bibnamefont
  {Leroux}}, \bibinfo {author} {\bibfnamefont {M.~H.}\ \bibnamefont
  {Schleier-Smith}}, \ and\ \bibinfo {author} {\bibfnamefont {V.}~\bibnamefont
  {Vuleti\ifmmode~\acute{c}\else \'{c}\fi{}}},\ }\href {\doibase
  10.1103/PhysRevLett.104.073602} {\bibfield  {journal} {\bibinfo  {journal}
  {Phys. Rev. Lett.}\ }\textbf {\bibinfo {volume} {104}},\ \bibinfo {pages}
  {073602} (\bibinfo {year} {2010})}\BibitemShut {NoStop}%
\bibitem [{\citenamefont {Sewell}\ \emph {et~al.}(2012)\citenamefont {Sewell},
  \citenamefont {Koschorreck}, \citenamefont {Napolitano}, \citenamefont
  {Dubost}, \citenamefont {Behbood},\ and\ \citenamefont
  {Mitchell}}]{Mitchell2012}%
  \BibitemOpen
  \bibfield  {author} {\bibinfo {author} {\bibfnamefont {R.~J.}\ \bibnamefont
  {Sewell}}, \bibinfo {author} {\bibfnamefont {M.}~\bibnamefont {Koschorreck}},
  \bibinfo {author} {\bibfnamefont {M.}~\bibnamefont {Napolitano}}, \bibinfo
  {author} {\bibfnamefont {B.}~\bibnamefont {Dubost}}, \bibinfo {author}
  {\bibfnamefont {N.}~\bibnamefont {Behbood}}, \ and\ \bibinfo {author}
  {\bibfnamefont {M.~W.}\ \bibnamefont {Mitchell}},\ }\href {\doibase
  10.1103/PhysRevLett.109.253605} {\bibfield  {journal} {\bibinfo  {journal}
  {Phys. Rev. Lett.}\ }\textbf {\bibinfo {volume} {109}},\ \bibinfo {pages}
  {253605} (\bibinfo {year} {2012})}\BibitemShut {NoStop}%
\bibitem [{\citenamefont {Hosten}\ \emph {et~al.}(2016)\citenamefont {Hosten},
  \citenamefont {Engelsen}, \citenamefont {Krishnakumar},\ and\ \citenamefont
  {Kasevich}}]{kasevich2016}%
  \BibitemOpen
  \bibfield  {author} {\bibinfo {author} {\bibfnamefont {O.}~\bibnamefont
  {Hosten}}, \bibinfo {author} {\bibfnamefont {N.~J.}\ \bibnamefont
  {Engelsen}}, \bibinfo {author} {\bibfnamefont {R.}~\bibnamefont
  {Krishnakumar}}, \ and\ \bibinfo {author} {\bibfnamefont {M.~A.}\
  \bibnamefont {Kasevich}},\ }\href {\doibase 10.1038/nature16176} {\bibfield
  {journal} {\bibinfo  {journal} {Nature}\ }\textbf {\bibinfo {volume} {529}},\
  \bibinfo {pages} {505} (\bibinfo {year} {2016})}\BibitemShut {NoStop}%
\bibitem [{\citenamefont {Cox}\ \emph {et~al.}(2016)\citenamefont {Cox},
  \citenamefont {Greve}, \citenamefont {Weiner},\ and\ \citenamefont
  {Thompson}}]{Cox2016}%
  \BibitemOpen
  \bibfield  {author} {\bibinfo {author} {\bibfnamefont {K.~C.}\ \bibnamefont
  {Cox}}, \bibinfo {author} {\bibfnamefont {G.~P.}\ \bibnamefont {Greve}},
  \bibinfo {author} {\bibfnamefont {J.~M.}\ \bibnamefont {Weiner}}, \ and\
  \bibinfo {author} {\bibfnamefont {J.~K.}\ \bibnamefont {Thompson}},\ }\href
  {\doibase 10.1103/PhysRevLett.116.093602} {\bibfield  {journal} {\bibinfo
  {journal} {Phys. Rev. Lett.}\ }\textbf {\bibinfo {volume} {116}},\ \bibinfo
  {pages} {093602} (\bibinfo {year} {2016})}\BibitemShut {NoStop}%
\bibitem [{\citenamefont {Christensen}\ \emph {et~al.}(2014)\citenamefont
  {Christensen}, \citenamefont {B\'eguin}, \citenamefont {Bookjans},
  \citenamefont {S\o{}rensen}, \citenamefont {M\"uller}, \citenamefont
  {Appel},\ and\ \citenamefont {Polzik}}]{Polzik2014}%
  \BibitemOpen
  \bibfield  {author} {\bibinfo {author} {\bibfnamefont {S.~L.}\ \bibnamefont
  {Christensen}}, \bibinfo {author} {\bibfnamefont {J.-B.}\ \bibnamefont
  {B\'eguin}}, \bibinfo {author} {\bibfnamefont {E.}~\bibnamefont {Bookjans}},
  \bibinfo {author} {\bibfnamefont {H.~L.}\ \bibnamefont {S\o{}rensen}},
  \bibinfo {author} {\bibfnamefont {J.~H.}\ \bibnamefont {M\"uller}}, \bibinfo
  {author} {\bibfnamefont {J.}~\bibnamefont {Appel}}, \ and\ \bibinfo {author}
  {\bibfnamefont {E.~S.}\ \bibnamefont {Polzik}},\ }\href {\doibase
  10.1103/PhysRevA.89.033801} {\bibfield  {journal} {\bibinfo  {journal} {Phys.
  Rev. A}\ }\textbf {\bibinfo {volume} {89}},\ \bibinfo {pages} {033801}
  (\bibinfo {year} {2014})}\BibitemShut {NoStop}%
\bibitem [{\citenamefont {Chen}\ \emph {et~al.}(2015)\citenamefont {Chen},
  \citenamefont {Hu}, \citenamefont {Duan}, \citenamefont {Braverman},
  \citenamefont {Zhang},\ and\ \citenamefont {Vuleti\ifmmode~\acute{c}\else
  \'{c}\fi{}}}]{Vuletic2015exp}%
  \BibitemOpen
  \bibfield  {author} {\bibinfo {author} {\bibfnamefont {W.}~\bibnamefont
  {Chen}}, \bibinfo {author} {\bibfnamefont {J.}~\bibnamefont {Hu}}, \bibinfo
  {author} {\bibfnamefont {Y.}~\bibnamefont {Duan}}, \bibinfo {author}
  {\bibfnamefont {B.}~\bibnamefont {Braverman}}, \bibinfo {author}
  {\bibfnamefont {H.}~\bibnamefont {Zhang}}, \ and\ \bibinfo {author}
  {\bibfnamefont {V.}~\bibnamefont {Vuleti\ifmmode~\acute{c}\else
  \'{c}\fi{}}},\ }\href {\doibase 10.1103/PhysRevLett.115.250502} {\bibfield
  {journal} {\bibinfo  {journal} {Phys. Rev. Lett.}\ }\textbf {\bibinfo
  {volume} {115}},\ \bibinfo {pages} {250502} (\bibinfo {year}
  {2015})}\BibitemShut {NoStop}%
\bibitem [{\citenamefont {McConnell}\ \emph {et~al.}(2013)\citenamefont
  {McConnell}, \citenamefont {Zhang}, \citenamefont {{\'{C} }uk}, \citenamefont
  {Hu}, \citenamefont {Schleier-Smith},\ and\ \citenamefont
  {Vuleti{\'{c}}}}]{vuletic_theory}%
  \BibitemOpen
  \bibfield  {author} {\bibinfo {author} {\bibfnamefont {R.}~\bibnamefont
  {McConnell}}, \bibinfo {author} {\bibfnamefont {H.}~\bibnamefont {Zhang}},
  \bibinfo {author} {\bibfnamefont {S.}~\bibnamefont {{\'{C} }uk}}, \bibinfo
  {author} {\bibfnamefont {J.}~\bibnamefont {Hu}}, \bibinfo {author}
  {\bibfnamefont {M.~H.}\ \bibnamefont {Schleier-Smith}}, \ and\ \bibinfo
  {author} {\bibfnamefont {V.}~\bibnamefont {Vuleti{\'{c}}}},\ }\href {\doibase
  10.1103/physreva.88.063802} {\bibfield  {journal} {\bibinfo  {journal}
  {Physical Review A}\ }\textbf {\bibinfo {volume} {88}},\ \bibinfo {pages}
  {063802} (\bibinfo {year} {2013})}\BibitemShut {NoStop}%
\bibitem [{\citenamefont {McConnell}\ \emph {et~al.}(2015)\citenamefont
  {McConnell}, \citenamefont {Zhang}, \citenamefont {Hu}, \citenamefont
  {{\'{C}}uk},\ and\ \citenamefont {Vuleti{\'{c}}}}]{vuletic_nature}%
  \BibitemOpen
  \bibfield  {author} {\bibinfo {author} {\bibfnamefont {R.}~\bibnamefont
  {McConnell}}, \bibinfo {author} {\bibfnamefont {H.}~\bibnamefont {Zhang}},
  \bibinfo {author} {\bibfnamefont {J.}~\bibnamefont {Hu}}, \bibinfo {author}
  {\bibfnamefont {S.}~\bibnamefont {{\'{C}}uk}}, \ and\ \bibinfo {author}
  {\bibfnamefont {V.}~\bibnamefont {Vuleti{\'{c}}}},\ }\href {\doibase
  10.1038/nature14293} {\bibfield  {journal} {\bibinfo  {journal} {Nature}\
  }\textbf {\bibinfo {volume} {519}},\ \bibinfo {pages} {439} (\bibinfo {year}
  {2015})}\BibitemShut {NoStop}%
\bibitem [{\citenamefont {Chase}\ and\ \citenamefont {Geremia}(2008)}]{chase}%
  \BibitemOpen
  \bibfield  {author} {\bibinfo {author} {\bibfnamefont {B.~A.}\ \bibnamefont
  {Chase}}\ and\ \bibinfo {author} {\bibfnamefont {J.~M.}\ \bibnamefont
  {Geremia}},\ }\href {\doibase 10.1103/PhysRevA.78.052101} {\bibfield
  {journal} {\bibinfo  {journal} {Phys. Rev. A}\ }\textbf {\bibinfo {volume}
  {78}},\ \bibinfo {pages} {052101} (\bibinfo {year} {2008})}\BibitemShut
  {NoStop}%
\bibitem [{\citenamefont {Zhang}\ \emph {et~al.}(2018)\citenamefont {Zhang},
  \citenamefont {Zhang},\ and\ \citenamefont {M{\o}lmer}}]{zhang}%
  \BibitemOpen
  \bibfield  {author} {\bibinfo {author} {\bibfnamefont {Y.}~\bibnamefont
  {Zhang}}, \bibinfo {author} {\bibfnamefont {Y.-X.}\ \bibnamefont {Zhang}}, \
  and\ \bibinfo {author} {\bibfnamefont {K.}~\bibnamefont {M{\o}lmer}},\ }\href
  {\doibase 10.1088/1367-2630/aaec36} {\bibfield  {journal} {\bibinfo
  {journal} {New Journal of Physics}\ }\textbf {\bibinfo {volume} {20}},\
  \bibinfo {pages} {112001} (\bibinfo {year} {2018})}\BibitemShut {NoStop}%
\bibitem [{\citenamefont {M{\o}lmer}\ and\ \citenamefont
  {Castin}(1996)}]{molmer_mcwf}%
  \BibitemOpen
  \bibfield  {author} {\bibinfo {author} {\bibfnamefont {K.}~\bibnamefont
  {M{\o}lmer}}\ and\ \bibinfo {author} {\bibfnamefont {Y.}~\bibnamefont
  {Castin}},\ }\href {\doibase 10.1088/1355-5111/8/1/007} {\bibfield  {journal}
  {\bibinfo  {journal} {Quantum Semiclass. Opt.}\ }\textbf {\bibinfo {volume}
  {8}},\ \bibinfo {pages} {49} (\bibinfo {year} {1996})}\BibitemShut {NoStop}%
\bibitem [{\citenamefont {Holstein}\ and\ \citenamefont
  {Primakoff}(1940)}]{PhysRev.58.1098}%
  \BibitemOpen
  \bibfield  {author} {\bibinfo {author} {\bibfnamefont {T.}~\bibnamefont
  {Holstein}}\ and\ \bibinfo {author} {\bibfnamefont {H.}~\bibnamefont
  {Primakoff}},\ }\href {\doibase 10.1103/PhysRev.58.1098} {\bibfield
  {journal} {\bibinfo  {journal} {Phys. Rev.}\ }\textbf {\bibinfo {volume}
  {58}},\ \bibinfo {pages} {1098} (\bibinfo {year} {1940})}\BibitemShut
  {NoStop}%
\bibitem [{\citenamefont {Deutsch}\ and\ \citenamefont
  {Jessen}(2010)}]{TheLongPaper}%
  \BibitemOpen
  \bibfield  {author} {\bibinfo {author} {\bibfnamefont {I.~H.}\ \bibnamefont
  {Deutsch}}\ and\ \bibinfo {author} {\bibfnamefont {P.~S.}\ \bibnamefont
  {Jessen}},\ }\href {\doibase https://doi.org/10.1016/j.optcom.2009.10.059}
  {\bibfield  {journal} {\bibinfo  {journal} {Optics Communications}\ }\textbf
  {\bibinfo {volume} {283}},\ \bibinfo {pages} {681} (\bibinfo {year}
  {2010})}\BibitemShut {NoStop}%
\bibitem [{Note1()}]{Note1}%
  \BibitemOpen
  \bibinfo {note} {For an initial spin coherent state, all spin-$1/2$ particles
  remain uncorrelated under incoherent optical pumping, hence the variance in
  $\protect \hat J_x$ and $\protect \hat J_y$ is the sum of the variances of
  all spins, which is constant in time.}\BibitemShut {Stop}%
\bibitem [{\citenamefont {Polkovnikov}(2010)}]{BoppRep2010}%
  \BibitemOpen
  \bibfield  {author} {\bibinfo {author} {\bibfnamefont {A.}~\bibnamefont
  {Polkovnikov}},\ }\href {\doibase https://doi.org/10.1016/j.aop.2010.02.006}
  {\bibfield  {journal} {\bibinfo  {journal} {Annals of Physics}\ }\textbf
  {\bibinfo {volume} {325}},\ \bibinfo {pages} {1790} (\bibinfo {year}
  {2010})}\BibitemShut {NoStop}%
\bibitem [{\citenamefont {Klimov}\ and\ \citenamefont
  {Romero}(2008)}]{gen_sphere_tensor}%
  \BibitemOpen
  \bibfield  {author} {\bibinfo {author} {\bibfnamefont {A.~B.}\ \bibnamefont
  {Klimov}}\ and\ \bibinfo {author} {\bibfnamefont {J.~L.}\ \bibnamefont
  {Romero}},\ }\href {\doibase 10.1088/1751-8113/41/5/055303} {\bibfield
  {journal} {\bibinfo  {journal} {Journal of Physics A: Mathematical and
  Theoretical}\ }\textbf {\bibinfo {volume} {41}},\ \bibinfo {pages} {055303}
  (\bibinfo {year} {2008})}\BibitemShut {NoStop}%
\bibitem [{\citenamefont {Sakurai}\ and\ \citenamefont
  {Napolitano}(2011)}]{Sakurai2011}%
  \BibitemOpen
  \bibfield  {author} {\bibinfo {author} {\bibfnamefont {J.~J.}\ \bibnamefont
  {Sakurai}}\ and\ \bibinfo {author} {\bibfnamefont {J.}~\bibnamefont
  {Napolitano}},\ }\href@noop {} {\emph {\bibinfo {title} {Modern quantum
  mechanics}}},\ \bibinfo {edition} {2nd}\ ed.\ (\bibinfo  {publisher} {San
  Francisco: Addison-Wesley},\ \bibinfo {year} {2011})\BibitemShut {NoStop}%
\bibitem [{\citenamefont {Toft}(1996)}]{Radon1996}%
  \BibitemOpen
  \bibfield  {author} {\bibinfo {author} {\bibfnamefont {P.~A.}\ \bibnamefont
  {Toft}},\ }\emph {\bibinfo {title} {The Radon Transform - Theory and
  Implementation}},\ \href {https://orbit.dtu.dk/files/5529668/Binder1.pdf}
  {Ph.D. thesis},\ \bibinfo  {school} {Technical University of Denmark}
  (\bibinfo {year} {1996})\BibitemShut {NoStop}%
\bibitem [{\citenamefont {Demkowicz-Dobrzanski}\ \emph
  {et~al.}(2012)\citenamefont {Demkowicz-Dobrzanski}, \citenamefont
  {Kolodynski},\ and\ \citenamefont {Guta}}]{DD}%
  \BibitemOpen
  \bibfield  {author} {\bibinfo {author} {\bibfnamefont {R.}~\bibnamefont
  {Demkowicz-Dobrzanski}}, \bibinfo {author} {\bibfnamefont {J.}~\bibnamefont
  {Kolodynski}}, \ and\ \bibinfo {author} {\bibfnamefont {M.}~\bibnamefont
  {Guta}},\ }\href {\doibase 10.1038/ncomms2067} {\bibfield  {journal}
  {\bibinfo  {journal} {Nature Communications}\ }\textbf {\bibinfo {volume}
  {3}},\ \bibinfo {pages} {1063} (\bibinfo {year} {2012})}\BibitemShut
  {NoStop}%
\bibitem [{\citenamefont {Braunstein}\ and\ \citenamefont
  {Caves}(1994)}]{Braunstein1994}%
  \BibitemOpen
  \bibfield  {author} {\bibinfo {author} {\bibfnamefont {S.~L.}\ \bibnamefont
  {Braunstein}}\ and\ \bibinfo {author} {\bibfnamefont {C.~M.}\ \bibnamefont
  {Caves}},\ }\href {\doibase 10.1103/PhysRevLett.72.3439} {\bibfield
  {journal} {\bibinfo  {journal} {Phys. Rev. Lett.}\ }\textbf {\bibinfo
  {volume} {72}},\ \bibinfo {pages} {3439} (\bibinfo {year}
  {1994})}\BibitemShut {NoStop}%
\bibitem [{\citenamefont {Paris}(2009)}]{QFI2009}%
  \BibitemOpen
  \bibfield  {author} {\bibinfo {author} {\bibfnamefont {M.~G.~A.}\
  \bibnamefont {Paris}},\ }\href {\doibase
  https://doi.org/10.1142/S0219749909004839} {\bibfield  {journal} {\bibinfo
  {journal} {Int. J. Quant. Inf.}\ }\textbf {\bibinfo {volume} {7}},\ \bibinfo
  {pages} {125} (\bibinfo {year} {2009})}\BibitemShut {NoStop}%
\bibitem [{\citenamefont {Kitagawa}\ and\ \citenamefont
  {Ueda}(1993)}]{Kitagawa1993}%
  \BibitemOpen
  \bibfield  {author} {\bibinfo {author} {\bibfnamefont {M.}~\bibnamefont
  {Kitagawa}}\ and\ \bibinfo {author} {\bibfnamefont {M.}~\bibnamefont
  {Ueda}},\ }\href {\doibase 10.1103/PhysRevA.47.5138} {\bibfield  {journal}
  {\bibinfo  {journal} {Phys. Rev. A}\ }\textbf {\bibinfo {volume} {47}},\
  \bibinfo {pages} {5138} (\bibinfo {year} {1993})}\BibitemShut {NoStop}%
\end{thebibliography}%

\end{document}